\documentclass[apj,iop]{emulateapj}
\pdfoutput=1
\usepackage{natbib}
\usepackage{longtable}
\usepackage{graphicx}
\usepackage[caption=false]{subfig}
\usepackage{float}
\usepackage{epstopdf}
\usepackage{amsmath}
\usepackage{appendix}
\usepackage{hyperref}

\newcommand{\HII}{H~{\footnotesize II}}

\newcommand{\msun}{$M_\odot$}

\shorttitle{X-ray and Radio Observations of BCDs}
\shortauthors{Latimer et al.}

\begin{document}

\title{An X-ray + Radio Search for Massive Black Holes in Blue Compact Dwarf Galaxies}

\author{Colin J. Latimer}
\affil{eXtreme Gravity Institute, Department of Physics, Montana State University, Bozeman, MT 59717, USA}
\email{colin.latimer@montana.edu}

\author{Amy E. Reines}
\affil{eXtreme Gravity Institute, Department of Physics, Montana State University, Bozeman, MT 59717, USA}

\author{Richard M. Plotkin}
\affil{International Centre for Radio Astronomy Research, Curtin University, GPO Box U1987, Perth, WA 6845, Australia}
\affil{Department of Physics, University of Nevada, Reno, NV 89557, USA}

\author{Thomas D. Russell}
\affil{Anton Pannekoek Institute for Astronomy, University of Amsterdam, Science Park 904, 1098 XH, Amsterdam, The Netherlands}

\and  

\author{James J. Condon}
\affil{National Radio Astronomy Observatory, Charlottesville, VA 22903, USA}

\begin{abstract}

Nearby blue compact dwarf (BCD) galaxies are arguably our best local analogues of galaxies in the earlier Universe that may host relics of black hole (BH) seeds. Here we present high-resolution {\it Chandra X-ray Observatory} and Karl G. Jansky Very Large Array (VLA) observations of five nearby BCDs with stellar masses of less than the Small Magellanic Cloud ($M_\star \sim 10^{7} - 10^{8.4}$ \msun). We search for signatures of accreting massive BHs at X-ray and radio wavelengths, which are more sensitive to lower BH accretion rates than optical searches. We detect a total of 10 hard X-ray sources and 10 compact radio sources at luminosities consistent with star-formation-related emission.  We find one case of a spatially-coincident X-ray and radio source within the astrometric uncertainties.  If the X-ray and radio emission are indeed coming from the same source, the origin of the radiation is plausibly from an active massive BH with log $(M_{\rm BH}/M_{\odot}) \sim 4.8 \pm 1.1$. However, given that the X-ray and radio emission are also coincident with a young star cluster complex, we consider the combination of an X-ray binary and a supernova remnant (or \HII\ region) a viable alternative explanation.  Overall, we do not find compelling evidence for active massive BHs in our target BCDs, which on average have stellar masses more than an order of magnitude lower than previous samples of dwarf galaxies found to host massive BHs. Our results suggest that moderately accreting massive BHs in BCDs are not so common as to permit unambiguous detection in a small sample.  

\end{abstract}

\section{Introduction}\label{sec:intro}

Over the past several years there has been mounting evidence for the existence of massive black holes (BHs) in at least some dwarf galaxies \citep[e.g.,][]{reines11,reines13,reines14,baldassare15,baldassare16,baldassare17,baldassare18, schramm13,moran14,lemons15,hainline16,pardo16,dickey19,nguyen19}. This has important implications for the formation and growth of the first ``seeds" of supermassive black holes that are ubiquitous in today's massive galaxies \citep[e.g.,][]{kormendyrichstone95,kormendyho13}. For example, present-day BH-host galaxy scaling relations at low masses and the BH occupation fraction in dwarf galaxies are predicted to be strong discriminants between seeds that were heavy ($\sim 10^5 M_\odot$) or light ($\sim 10^2 M_\odot$) \citep[e.g.,][]{volonteri10,greene12,ricarte18}.

While there has been recent progress on the observational front, the BH occupation fraction in dwarf galaxies is not yet well determined (although see \citealt{milleretal2015}) and we know very little about the types of dwarf galaxies that may be preferential hosts to massive BHs. Dwarf galaxies come in a variety of flavors, including star-forming irregulars, spirals and blue compact dwarfs, early-type ellipticals and ultracompact dwarfs, and dark-matter-dominated dwarf spheroidals and ultra-faint dwarfs.  So far the largest samples of massive BHs in dwarf galaxies have been found by applying optical emission line diagnostics (e.g., BPT selection and broad H$\alpha$; \citealt{reines13}) to Sloan Digital Sky Survey (SDSS) spectroscopy.  However, these types of optical searches are biased towards active BHs that are more easily observed, such as those with high Eddington ratios and those that are unobscured by star formation.  Indeed, the host galaxies of the \citet{reines13} sample tend to have bright nuclear point sources (from the active galactic nuclei, i.e., AGN), regular morphologies, and little ongoing star formation (Schutte et al., in prep; Kimbrell et al., in prep).  This is similar to the slightly more massive counterparts in the earlier sample of \cite{greeneho07}.

Given that our understanding of the demographics of massive BHs in different categories of dwarf galaxies is quite limited, and the surprising discovery of a massive BH in the blue compact dwarf (BCD) galaxy Henize 2-10 \citep{reines11}, we have observed a small sample of BCDs using X-ray and radio observations. High-resolution observations from the {\it Chandra X-ray Observatory} and the Very Large Array (VLA) are especially well-suited to searching for weakly accreting massive BHs and those residing in star-forming host galaxies, as exemplified by detailed studies of Henize 2-10 \citep{reines11,reines12,reines16} that reveal a spatially coincident compact, non-thermal radio and X-ray point source at the center of the galaxy.\footnote{Although see \cite{hebbar19} who argue for a supernova remnant origin.  We defer a detailed discussion of this object to a forthcoming paper.}
Moreover, BCDs are characterized by blue colors, low metallicities, and concentrated regions of intense star formation \citep[e.g.][]{gil03}, making them our best local analogues of low-mass, physically small, gas-rich, star-forming galaxies in the earlier Universe, where the first BH seeds likely formed.

\section{Sample of Blue Compact Dwarf Galaxies} \label{sec:samp}

Our target galaxies are taken from the sample of BCDs from \citet{gil03}. \citet{gil03} propose that a galaxy must fulfill the following observational criteria in order to be classified as a BCD: (1) blue with $\mu_{B,{\rm peak}} - \mu_{R,{\rm peak}} \lesssim 1$ mag arcsec$^{-2}$, (2) compact with $\mu_{B,{\rm peak}} < 22$ mag arcsec$^{-2}$, and (3) a dwarf with $M_K > -21$ mag.  This definition separates the BCDs from other types of dwarf galaxies such as dwarf irregulars and dwarf ellipticals. Given that our search technique depends on high-resolution X-ray and radio observations, we adopted two complementary strategies for selecting our target BCDs from \cite{gil03}.  

First, we selected galaxies detected at 1.4 GHz in the {Faint Images of the Radio Sky at Twenty-cm} ({FIRST}) survey \citep{becker95} with distances $< 20$ Mpc. BCDs detected by {FIRST} are preferred targets since they must either have an AGN and/or intense star formation to produce the observed radio emission on scales of ${\sim} 5\arcsec$, making them good analogues of Henize~2-10. The distance cut was applied to retain both sensitivity to low luminosities and fine enough spatial resolution to accurately determine if X-ray and radio sources are co-incident. Out of the 80 BCDs in \cite{gil03} covered by the {FIRST} survey area, four galaxies met these selection criteria: Haro 3, Haro 9, Mrk 709\footnote{The distance of 15.7 Mpc for Mrk 709 provided in \cite{gil03} (from NED) was incorrect. \citet{reines14} present a redshift of $z=0.052~(d \sim 214$ Mpc) based on SDSS spectroscopy.} and II Zw 70.  These four BCDs were the subject of a joint {\it Chandra}--VLA proposal (PI: Reines; CXO proposal number 13700563, VLA project SD0563).  Our study of Mrk 709, and its candidate massive BH, has previously been presented in \citet{reines14}. 

To expand our sample, we also searched for BCDs in the sample of \cite{gil03} with available observations in the {\it Chandra} archive.  
We found 18 galaxies with {\it Chandra} coverage and, after obtaining and inspecting these data, found five BCDs with detectable hard X-ray point sources worthy of radio follow-up with the VLA (any existing VLA archival data were not deep enough and/or lacked sufficient resolution to meet our requirements). We proposed for VLA observations of these preferred targets with existing X-ray sources, however only two (Mrk 996 and SBS 0940+544) were ultimately observed for this project (PI Reines; VLA project 12B-206) due to scheduling priorities. We did not propose for the remaining 13 galaxies without detectable {\it Chandra} point sources since we are primarily interested in finding spatially coincident X-ray and radio sources as a signpost of massive BH accretion.



\begin{figure*}
\centering
\subfloat[II Zw 70 \label{fig:sdssa}]{\includegraphics[width=0.18\textwidth]{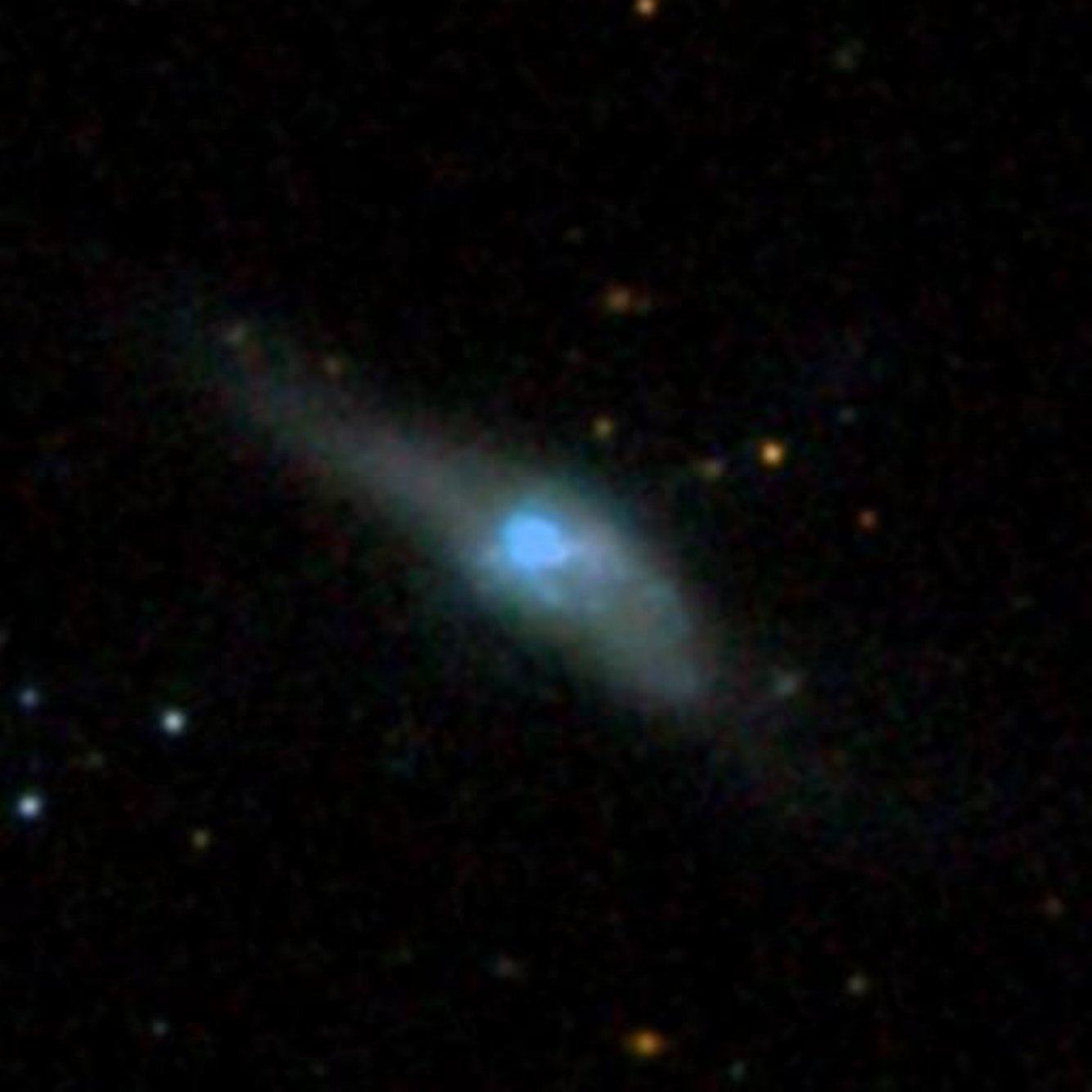}}\hfill
\subfloat[Haro 3 \label{fig:sdssb}]{\includegraphics[width=0.18\textwidth]{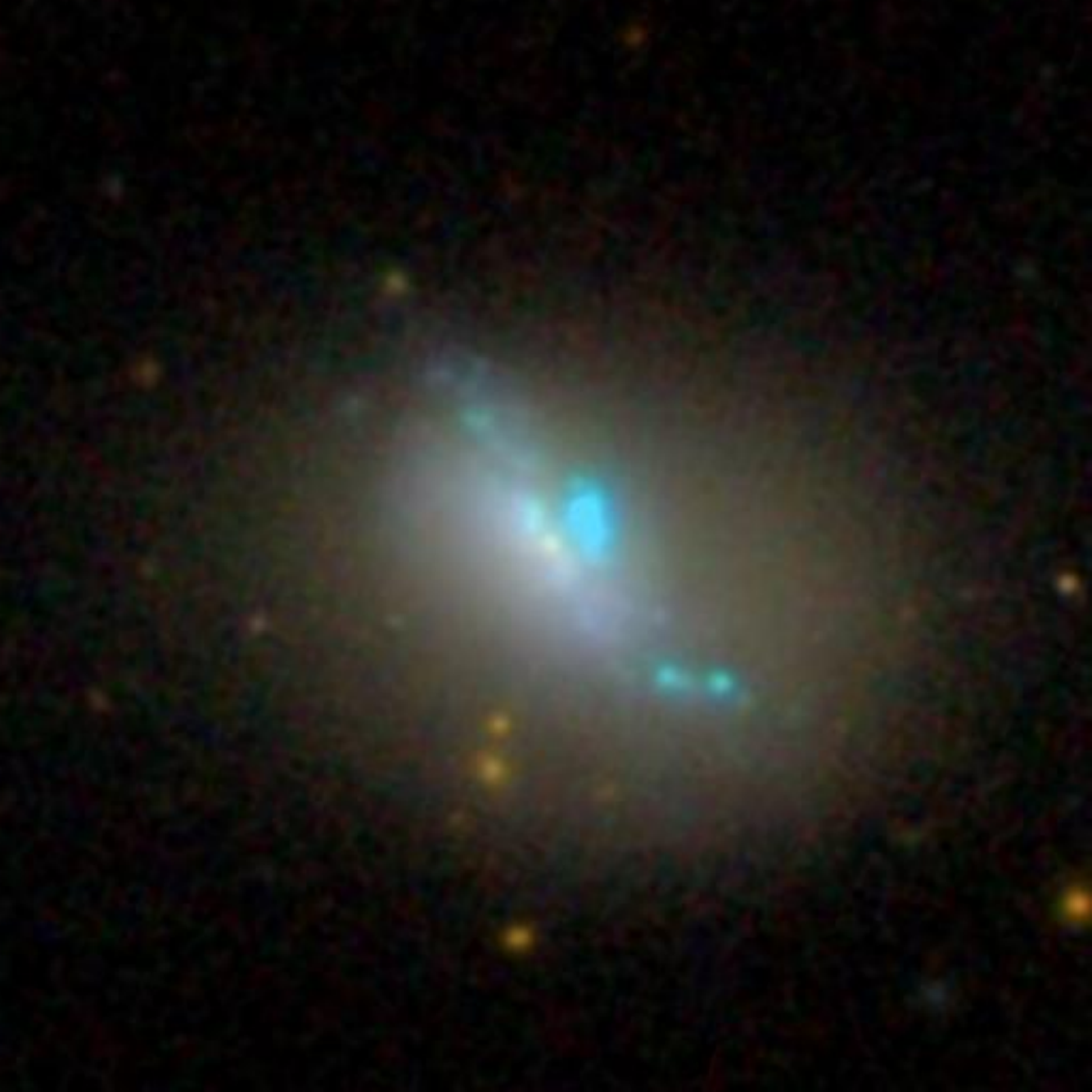}}\hfill
\subfloat[Haro 9 \label{fig:sdssc}]{\includegraphics[width=0.18\textwidth]{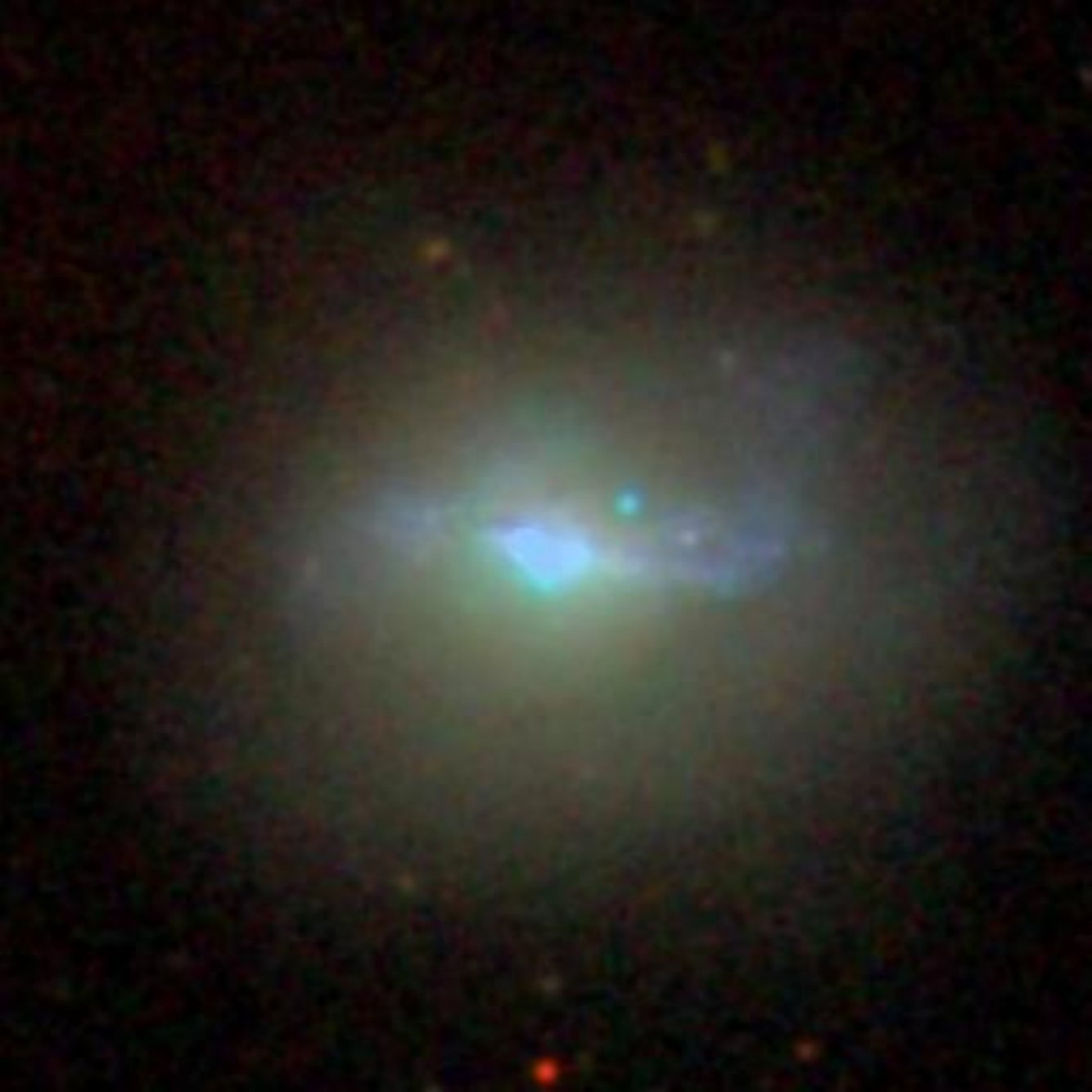}}\hfill
\subfloat[Mrk 996 \label{fig:sdssd}]{\includegraphics[width=0.18\textwidth]{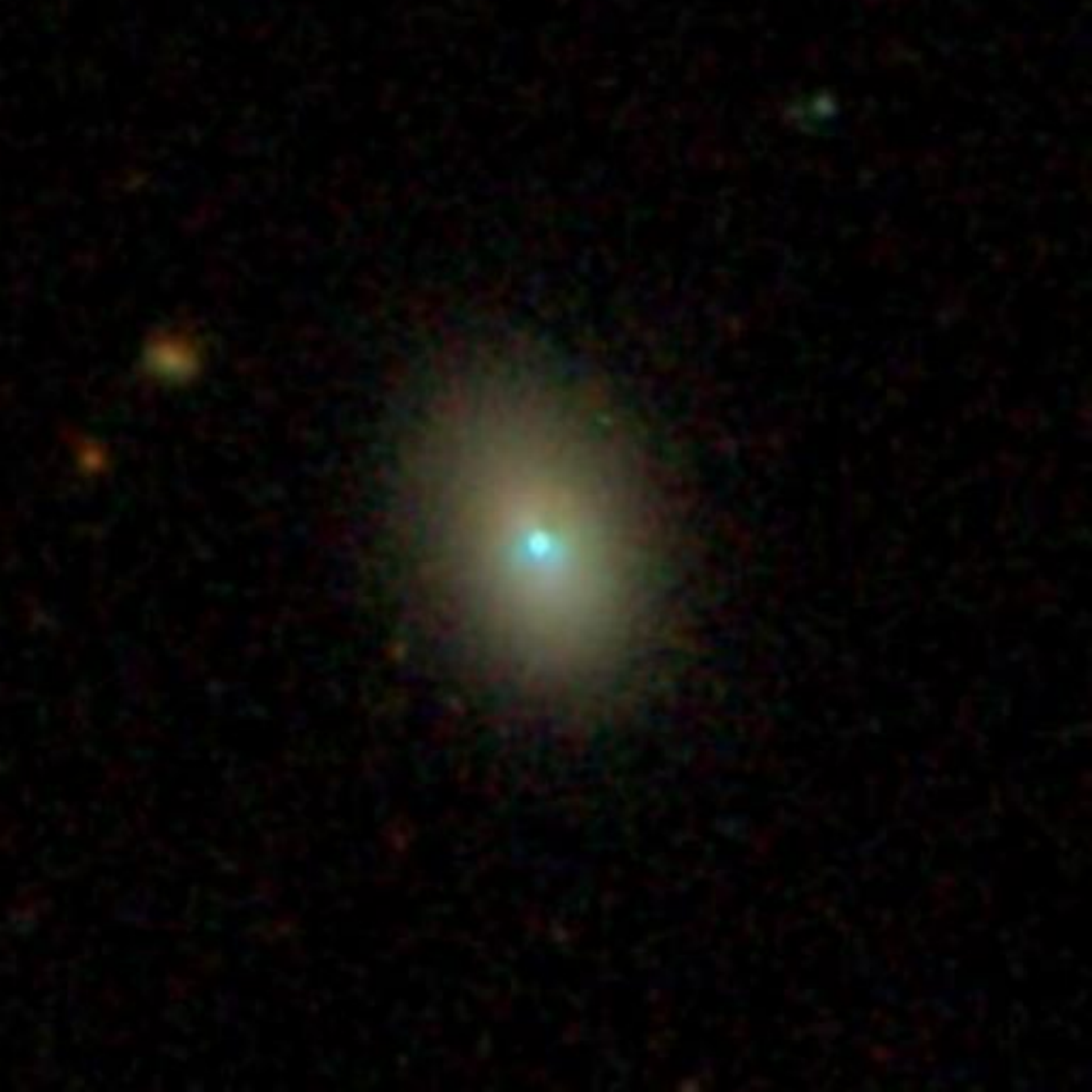}}\hfill
\subfloat[SBS 0940+544 \label{fig:sdsse}]{\includegraphics[width=0.18\textwidth]{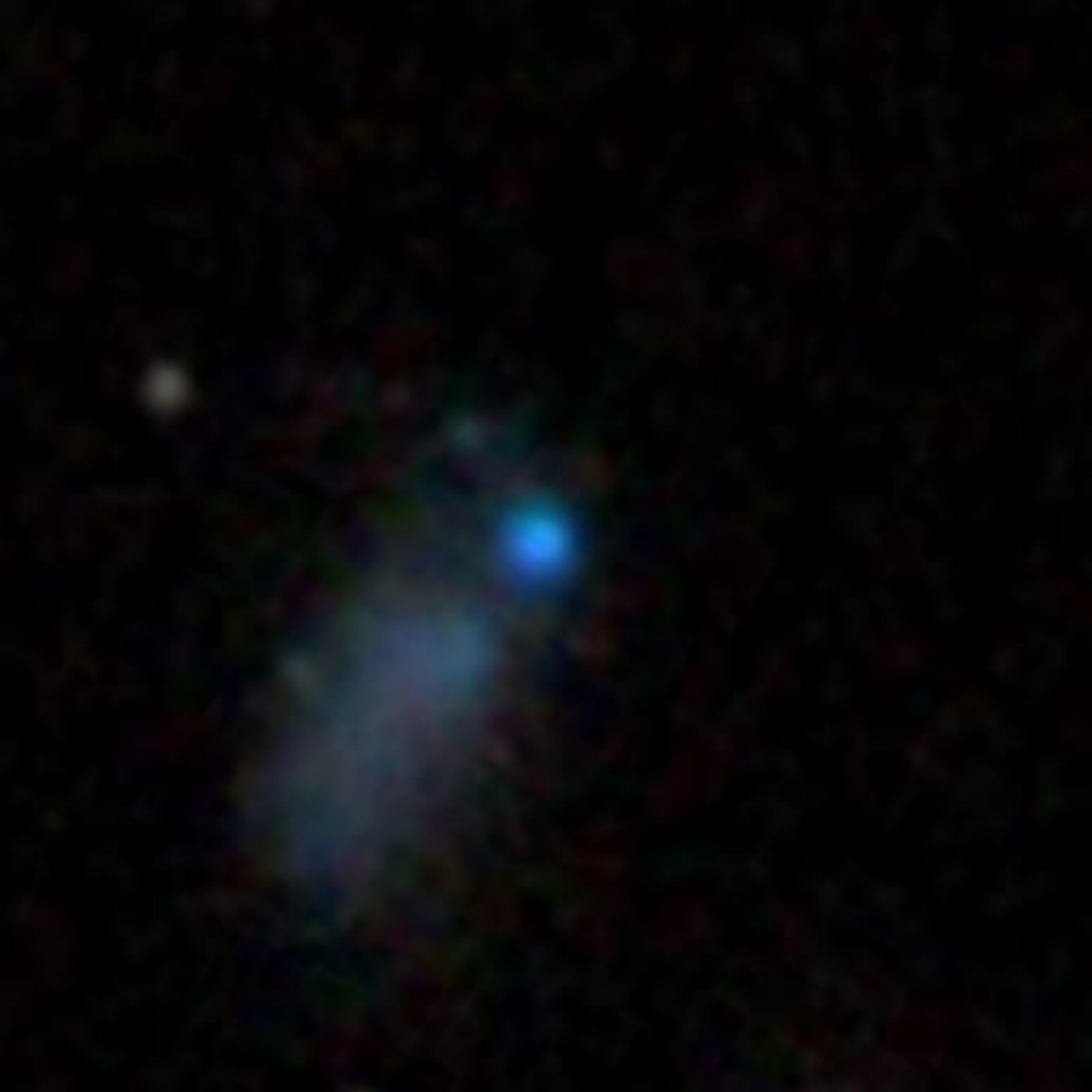}}
\caption{SDSS color composite images of the galaxies in our sample. Each image has dimensions of $1.69' \times 1.69'$, except for (e), which has dimensions of $0\farcm85 \times 0\farcm85$.} 
\label{fig:sdss}
\end{figure*}

Our final sample consists of five BCD galaxies (see Figure \ref{fig:sdss}) and their properties are given in Table \ref{tab:sample}. Note that while distances from \cite{gil03} were used to initially select our sample, in our analysis here we use distances derived from redshifts in the NASA-Sloan Atlas\footnote{\url{http://nsatlas.org}, version 0.1.2} (NSA) with $h=0.73$.

\begin{deluxetable*}{ccccccccrc}
\tabletypesize{\footnotesize}
\tablecaption{Sample of Blue Compact Dwarf Galaxies}
\tablewidth{0pt}
\tablehead{
\colhead{Galaxy} & \colhead{R.A.} & \colhead{Decl.}  & \colhead{$N_{\rm H}$} & \colhead{$z$} & \colhead{$r_{50}$} & \colhead{Distance} & \colhead{$M_g$} & \colhead{$g-i$} & \colhead{log $M_\star$/\msun}  \\
\colhead{ } & \colhead{ } & \colhead{ } & \colhead{($10^{20}$ cm$^{-2}$)} & \colhead{ } & \colhead{(kpc)}  & \colhead{(Mpc)} & \colhead{(mag)} & \colhead{(mag)} & \colhead{ } \\
\colhead{(1)} & \colhead{(2)} & \colhead{(3)} & \colhead{(4)} & \colhead{(5)} & \colhead{(6)}  & \colhead{(7)} & \colhead{(8)} & \colhead{(9)} & \colhead{(10)} }
\startdata
II Zw 70  		 &  14 50 56.5  & +35 34 18     & $1.16$  & 0.00470 & 0.30 & 19.31  & $-16.98$ & $-0.289$ & 7.28 \\
Haro 3    		 &  10 45 22.4  & +55 57 37     & $0.66$  & 0.00409 & 0.61 & 16.81  & $-18.21$ & $-0.062$ & 8.03 \\
Haro 9    		 &  12 45 17.1  & +27 07 32     & $0.71$  & 0.00443 & 0.84 & 18.18  & $-18.92$ &    0.103    & 8.38 \\
Mrk 996 		 &  01 27 35.5  & $-06$ 19 36 &  $3.83$ & 0.00491 & 0.46 & 20.18  & $-17.38$ &    0.081    & 7.73 \\
SBS 0940+544 &  09 44 16.6 & +54 11 34      & $1.34$  & 0.00652 & 0.83 & 26.78  & $-15.97$ & $-0.256$ & 7.04 
\enddata
\tablecomments{Column 1: galaxy name.  
Column 2: right ascension in units of hours, minutes, seconds (J2000).  
Column 3: declination in units of degrees, arcminutes, arcseconds (J2000). 
Column 4: galactic neutral hydrogen column density.  
Column 5: redshift, specifically the \texttt{zdist} parameter from the NSA.
Column 6: Petrosian 50\% light radius.  
Column 7: distance estimate from \texttt{zdist}.
Column 8: absolute g-band magnitude corrected for foreground Galactic extinction.
Column 9: $g-i$ color.  
Column 10: log galaxy stellar mass.
The values given in columns 5-10 are from the NSA\footnote{SBS 0940+544 appears as two separate sources in the NSA; one for the bright central region, and one for the trailing tail region (see Figure \ref{fig:sdss}). The values for $M_g$, $g-i$ color, and log $M_*$ (Columns 8, 9, and 10) are from the combined sources, while the values for $z$, $r_{50}$, and distance (Columns 5, 6, and 7) are from the bright central region 
(though the trailing tail region has similar values).} and we assume $h=0.73$. }
\label{tab:sample}
\end{deluxetable*}

\section{Observations and Data Reduction} \label{sec:obstot}

\subsection{Chandra X-ray Observatory} \label{sec:xrayobs}

X-ray observations of our target galaxies were taken with \textit{Chandra} between 2009 Sep 03 and 2012 Dec 31.  Exposure times were between 16.8 ks and 53.3 ks.  A summary of the {\it Chandra} X-ray observations is given in Table \ref{tab:cxo}.

Each dwarf galaxy was placed at the aimpoint of the ACIS S3 chip.  We  used the \textit{Chandra} Interactive Analysis of Observations ({\tt CIAO}) software v4.7 \citep{fruscione06} to reduce each observation.  We first reprocessed the data by applying calibration files (CALDB 4.6.7), and we determined that no observations were affected by background flares.  

We then aligned each \textit{Chandra} observation to the optical SDSS astrometric frame, using the {\tt CIAO} tool {\tt reproject\_aspect}.  To align the astrometry, we created a list of X-ray point sources on the S3 chip by running the point source detection algorithm {\tt wavdetect} on a \textit{Chandra} image filtered from 0.5-7~keV.  We then excluded all X-ray sources falling within 3$r_{50}$ of the dwarf galaxy, and we correlated the remaining X-ray sources to optical point sources in the SDSS DR12 with $i<22$ mag.  We found 2-5 common X-ray and optical point sources on the S3 chip for each observation.  Given the small number of common sources, we only applied a translation correction to the \textit{Chandra} images in the x,y directions.  The applied astrometric shifts range from $\pm$0.01-1.3 pixels, and the median shifts across all five observations were $\left|\Delta x\right|=0.2$ and $\left|\Delta y\right|=0.4$ pixels ($0\farcs1$ and $0\farcs2$, respectively).  

\begin{deluxetable*}{ccccc}
\tabletypesize{\footnotesize}
\tablecaption{Chandra Observations}
\tablewidth{0pt}
\tablehead{
\colhead{Galaxy}  & \colhead{Date observed}  & \colhead{Obs ID} & \colhead{Exp.\ time (ks)} & \colhead{$N_{\rm background}$} }
\startdata
II Zw 70   & 2011 Nov 06 & 13930 & 30.6 & 0.033 \\
Haro 3   & 2012 Dec 31 & 13927 & 17.8  &  0.131 \\
Haro 9  & 2012 Nov 13 & 13928  & 16.8 &  0.176 \\
Mrk 996  & 2009 Sep 03 & 11567 & 53.3 &  0.104 \\
SBS 0940+544  & 2010 Jan 18 & 11288 & 16.8 & 0.079 
\enddata
\tablecomments{$N_{\rm background}$ is the number of expected 2-10 keV $N(>S)$ background sources within $3r_{50}$, using \citet{moretti03}. }
\label{tab:cxo}
\end{deluxetable*}

\subsection{Karl G. Jansky Very Large Array} \label{sec:vlaobs}

Observations of our sample of galaxies were taken with the VLA in its extended A-configuration between 2012 December 1 and 2013 January 1. The observations typically provided between 50 minutes and 90 minutes of on source time (see Table \ref{tab:vla}). All observations were taken at $C$-band with two 1-GHz wide basebands centered at 5.0 and 7.4 GHz, each comprised of eight 128-MHz sub-bands containing 64 spectral channels of width 2-MHz. The primary (bandpass and amplitude) and secondary (phase) calibrators are given in Table \ref{tab:vla}.

The data were edited and calibrated following standard procedures within the Common Astronomy Software Application \citep[CASA;][]{mcmullin07}. Using natural weighting for maximum sensitivity, we imaged the calibrated data separately at 5.0 and 7.4 GHz (Figure \ref{fig:radio}), where deconvolution (cleaning) was carried out with the multi-frequency synthesis algorithm within CASA. The RMS noise and beam parameters are given for each image in Table \ref{tab:vlaim}. We also explored different antenna weightings, and used UV-cuts to filter out diffuse emission (as well as match spatial sensitivity between the 5.0 and 7.4 GHz images). However, these did not alter the results, and, therefore, we only present the images using natural weighting to maximize the sensitivity. The absolute astrometry of the VLA observations is accurate to $\lesssim 0\farcs1$.

\begin{deluxetable*}{ccccccc}
\tabletypesize{\footnotesize}
\tablecaption{VLA Observations}
\tablewidth{0pt}
\tablehead{
\colhead{Galaxy}  & \colhead{Date Observed}  & \colhead{Project Code} & Flux Calibrator & Phase Calibrator & \colhead{Time on Source (min)}}
\startdata
II Zw 70  & 2012 Dec 01 & SD0563 & 3C 286 & J1416+3444 & 90   \\
Haro 3  & 2012 Dec 02 & SD0563 & 3C 286 & J1035+5628 & 50    \\
Haro 9   & 2012 Dec 01 & SD0563 & 3C 286 & J1221+2813 & 80   \\
Mrk 996  & 2013 Jan 01 & 12B-206  & 3C 48 & J0110$-$0741 & 90   \\
SBS 0940+544  & 2012 Dec 24 & 12B-206 & 3C 147 & J0932+5306 & 50  
\enddata
\tablecomments{All observations were taken at $C$-band while the VLA was in the A-configuration.}
\label{tab:vla}
\end{deluxetable*}
\begin{figure*}
\centering
\subfloat[II Zw 70 VLA 5 GHz image with contour levels of $(\sqrt{2})^n$ times the RMS noise, where $n=5$, 7, and 9. \label{fig:radioa}]{\includegraphics[width=0.48\textwidth]{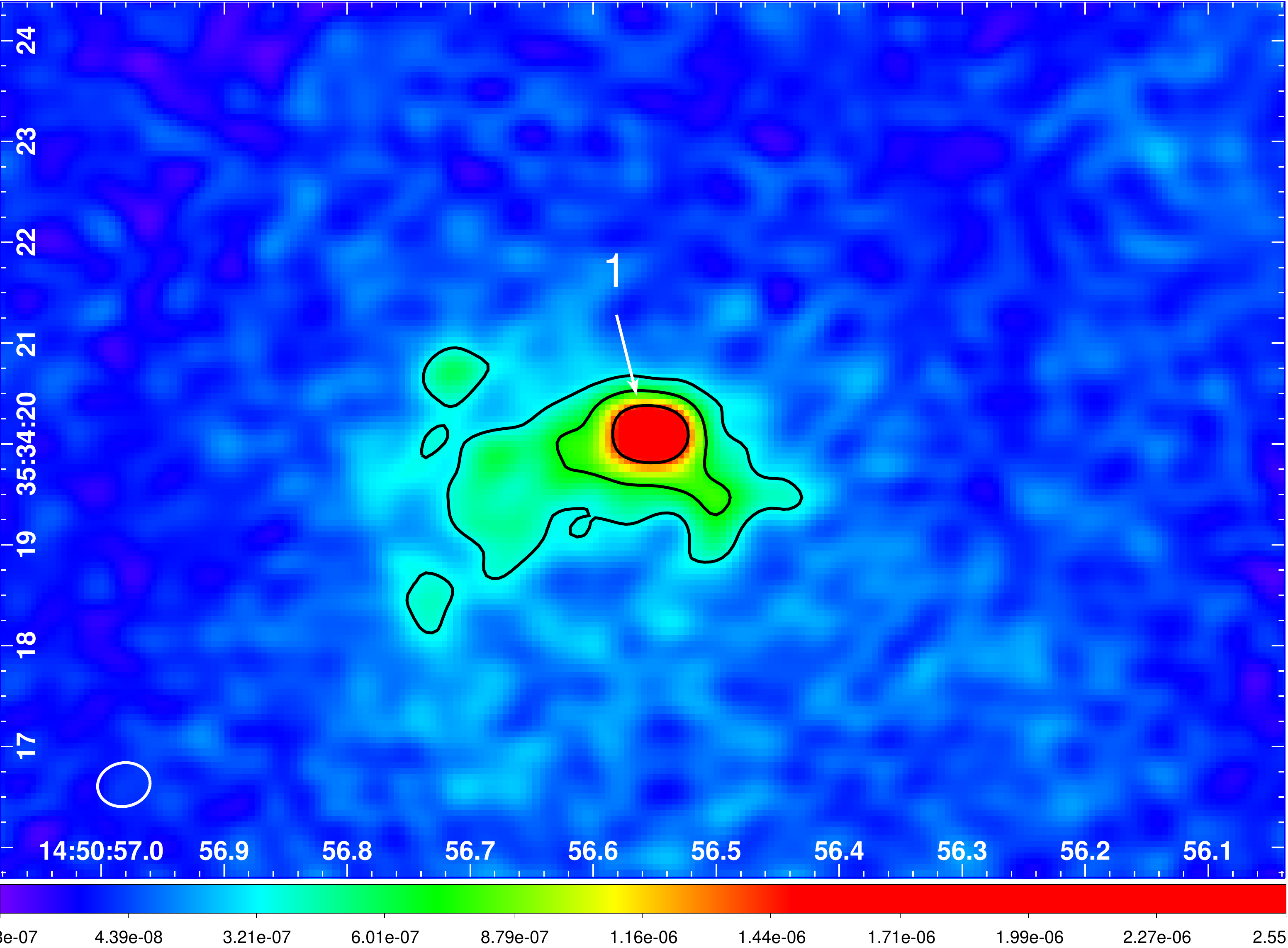}}\hfill
\subfloat[II Zw 70 VLA 7.4 GHz image with the 5 GHz contour levels overlaid.  \label{fig:radiob}]{\includegraphics[width=0.48\textwidth]{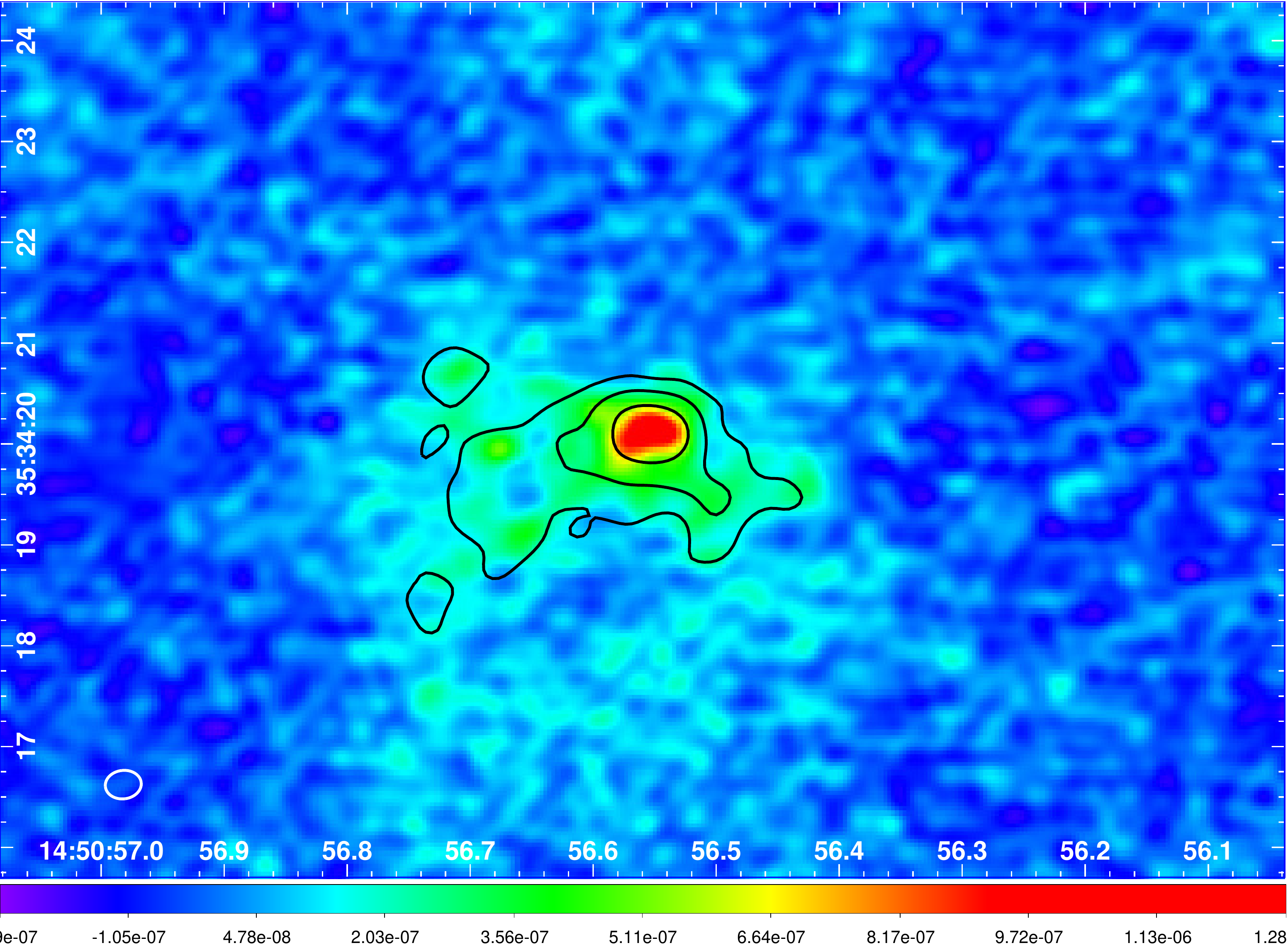}}\hfill
\subfloat[Haro 3 VLA 5 GHz image with contour levels of $(\sqrt{2})^n$ times the RMS noise, where $n=4$, 5, 7, 9, and 10. \label{fig:radioc}]{\includegraphics[width=0.48\textwidth]{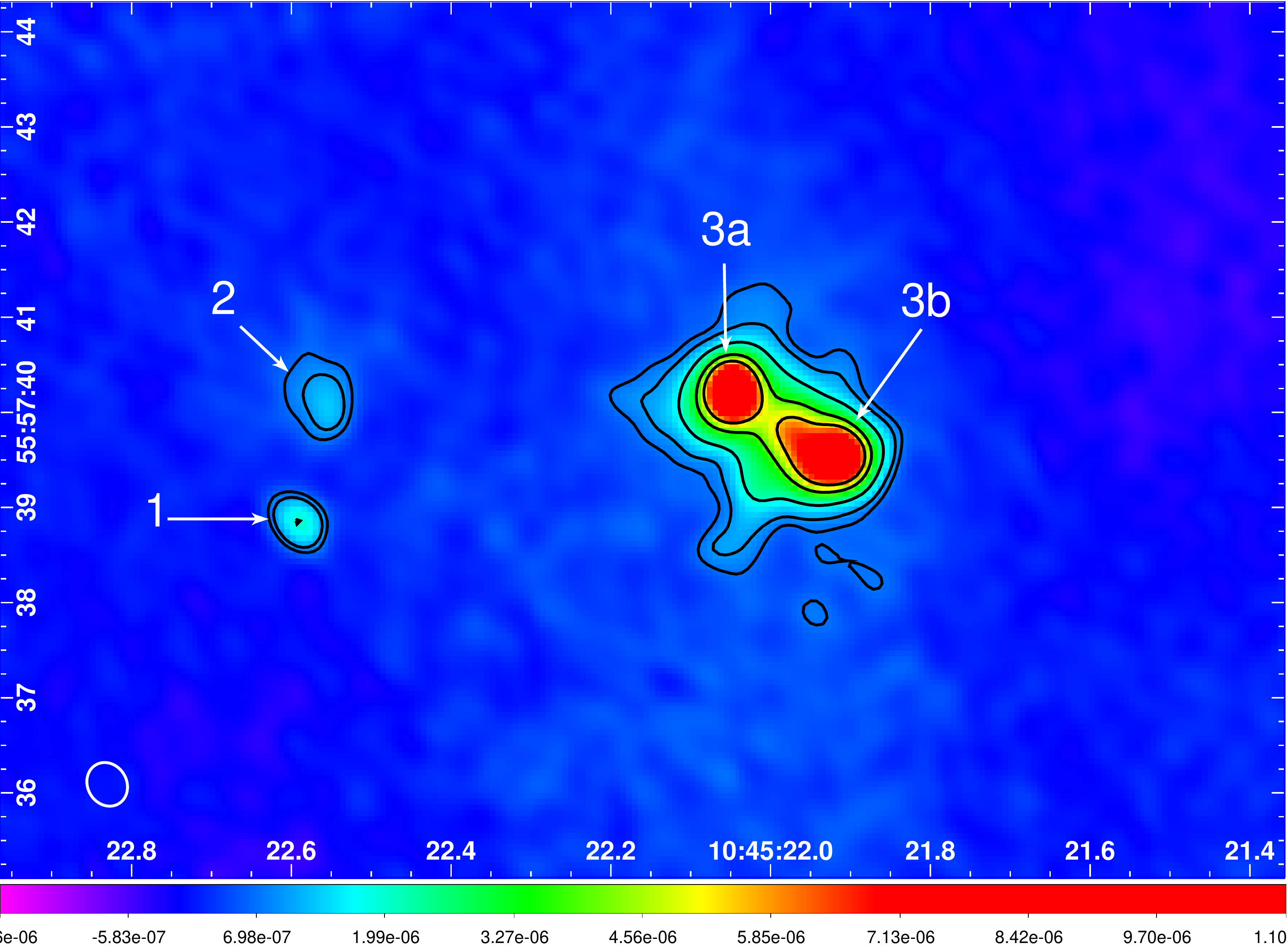}}\hfill
\subfloat[Haro 3 VLA 7.4 GHz image with the 5 GHz contour levels overlaid..  \label{fig:radiod}]{\includegraphics[width=0.48\textwidth]{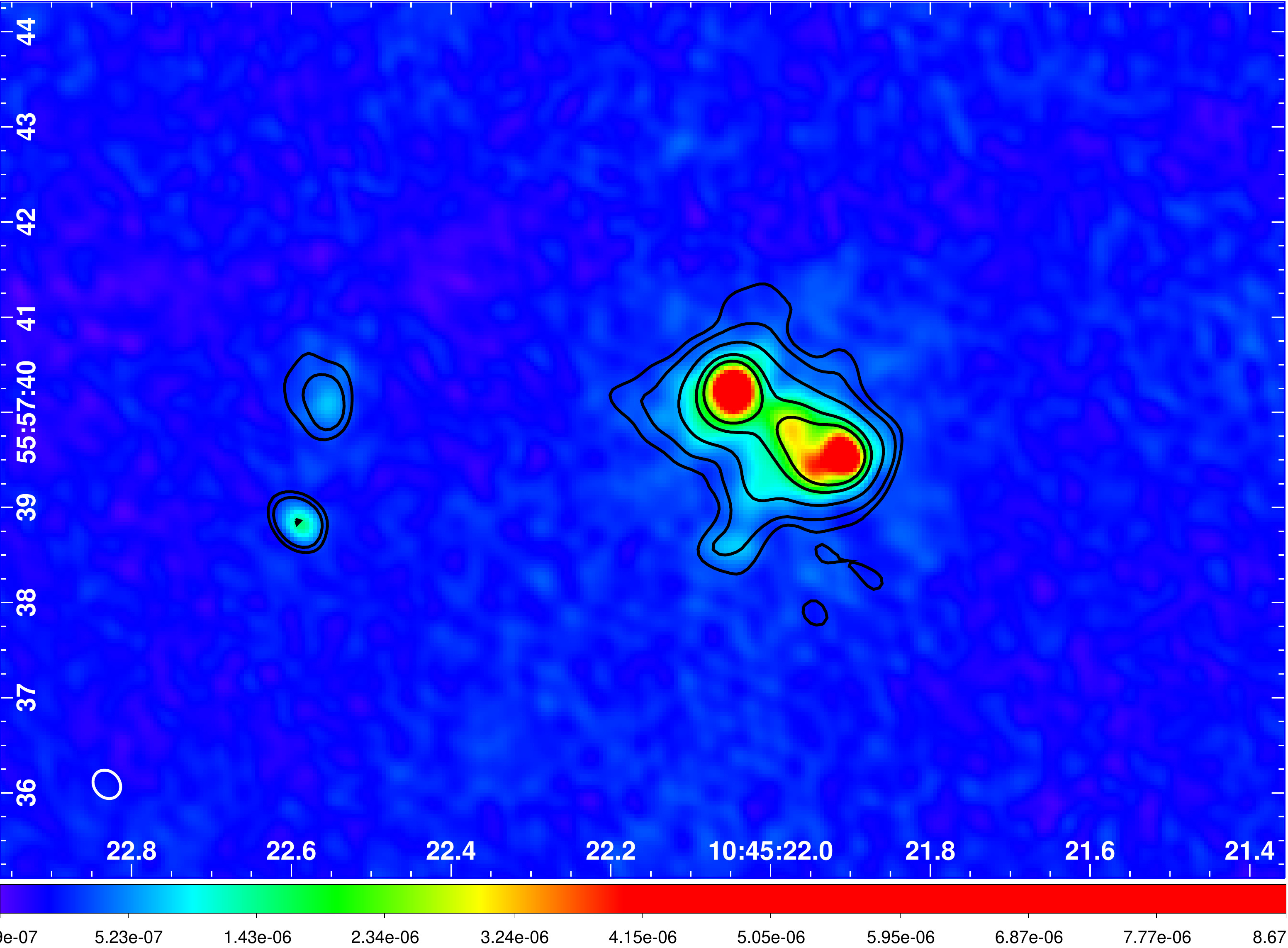}}\hfill
\subfloat[Haro 9 VLA 5 GHz image with contour levels of $(\sqrt{2})^n$ times the RMS noise, where $n=4$, 5, and 6. \label{fig:radioe}]{\includegraphics[width=0.48\textwidth]{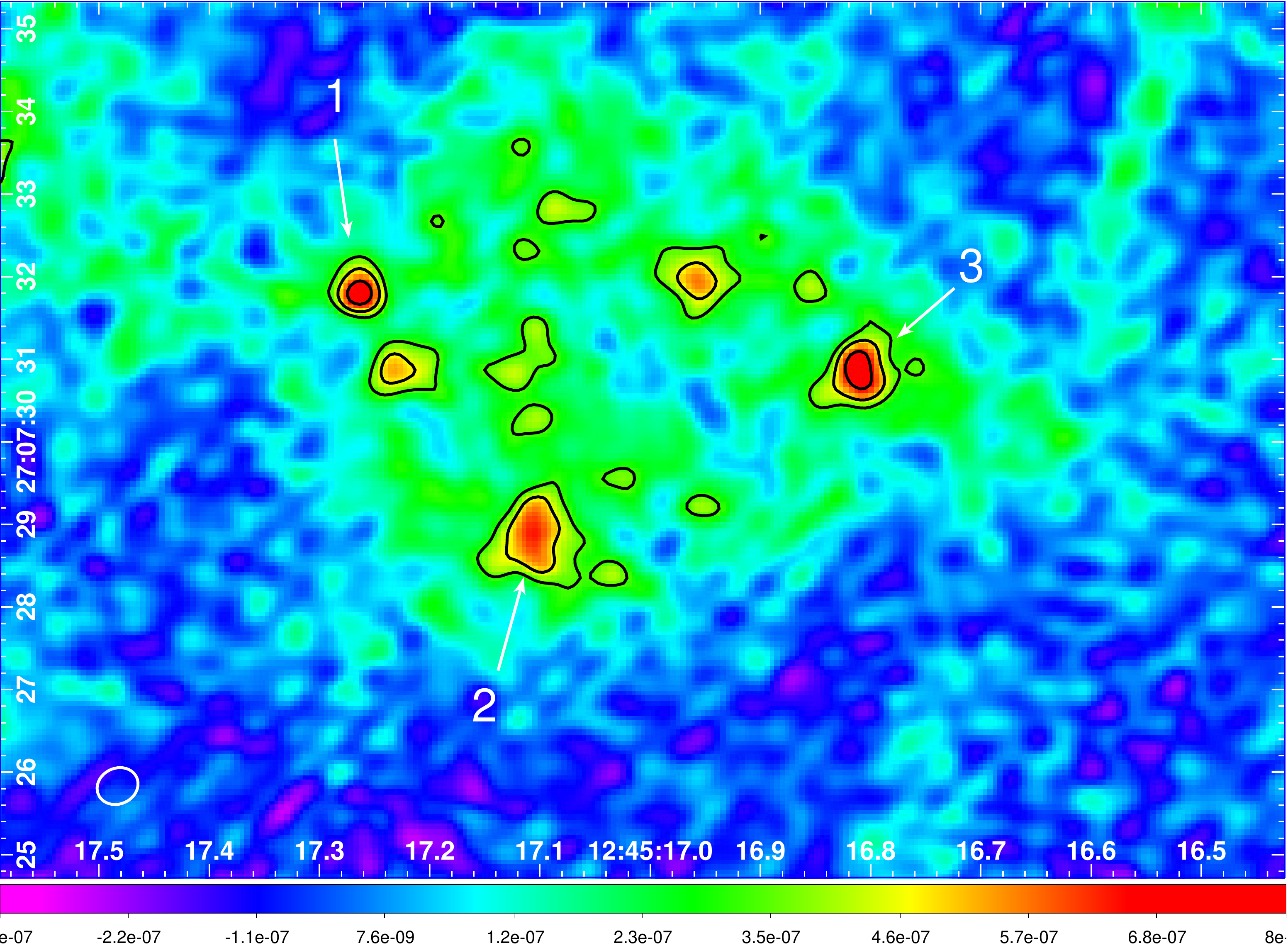}}\hfill
\subfloat[Haro 9 VLA 7.4 GHz image with the 5 GHz contour levels overlaid.  \label{fig:radiof}]
{\includegraphics[width = 0.48\textwidth]{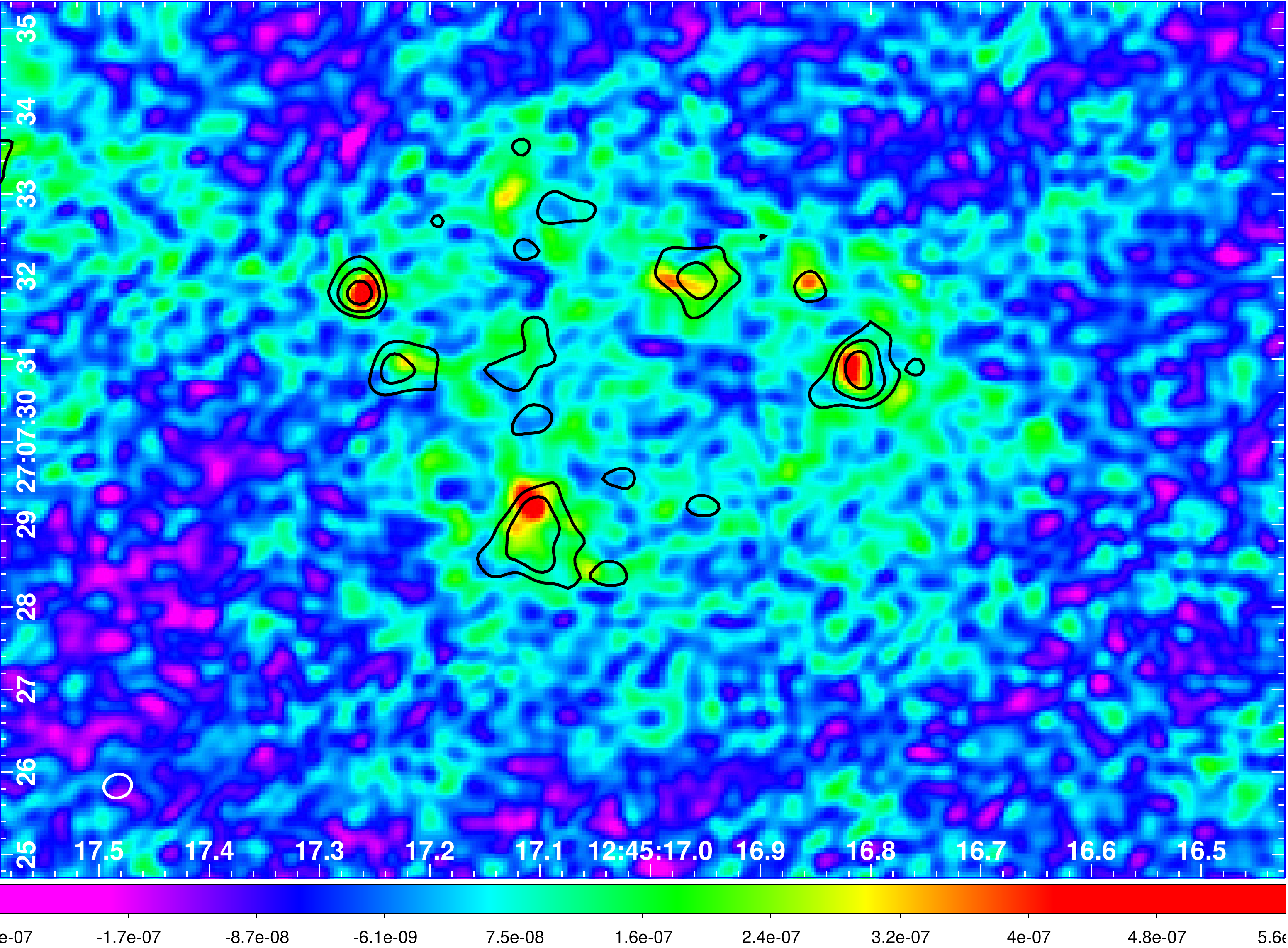}}
\caption{VLA 5 GHz (left) and 7.4 GHz (right) images of the galaxies in our sample. The beam sizes are shown in the lower left corners.}
\label{fig:radio}
\end{figure*}
\begin{figure*}
\ContinuedFloat
\subfloat[Mrk 996 VLA 5 GHz image with contour levels of $(\sqrt{2})^n$ times the RMS noise, where $n=3$, 5, 8, and 10. \label{fig:radiog}]{\includegraphics[width=0.48\textwidth]{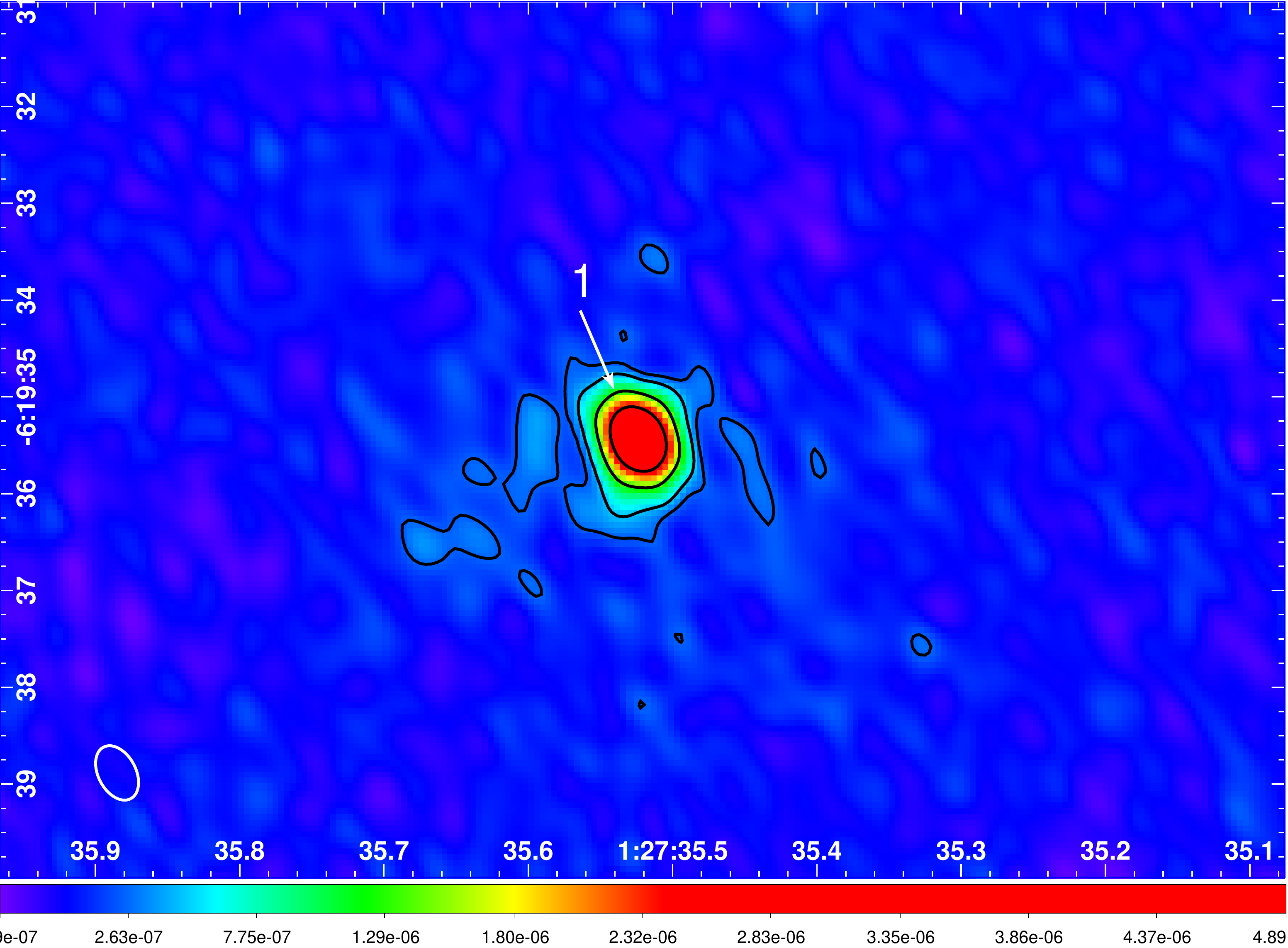}}\hfill
\subfloat[Mrk 996 VLA 7.4 GHz image with the 5 GHz contour levels overlaid. \label{fig:radioh}]{\includegraphics[width=0.48\textwidth]{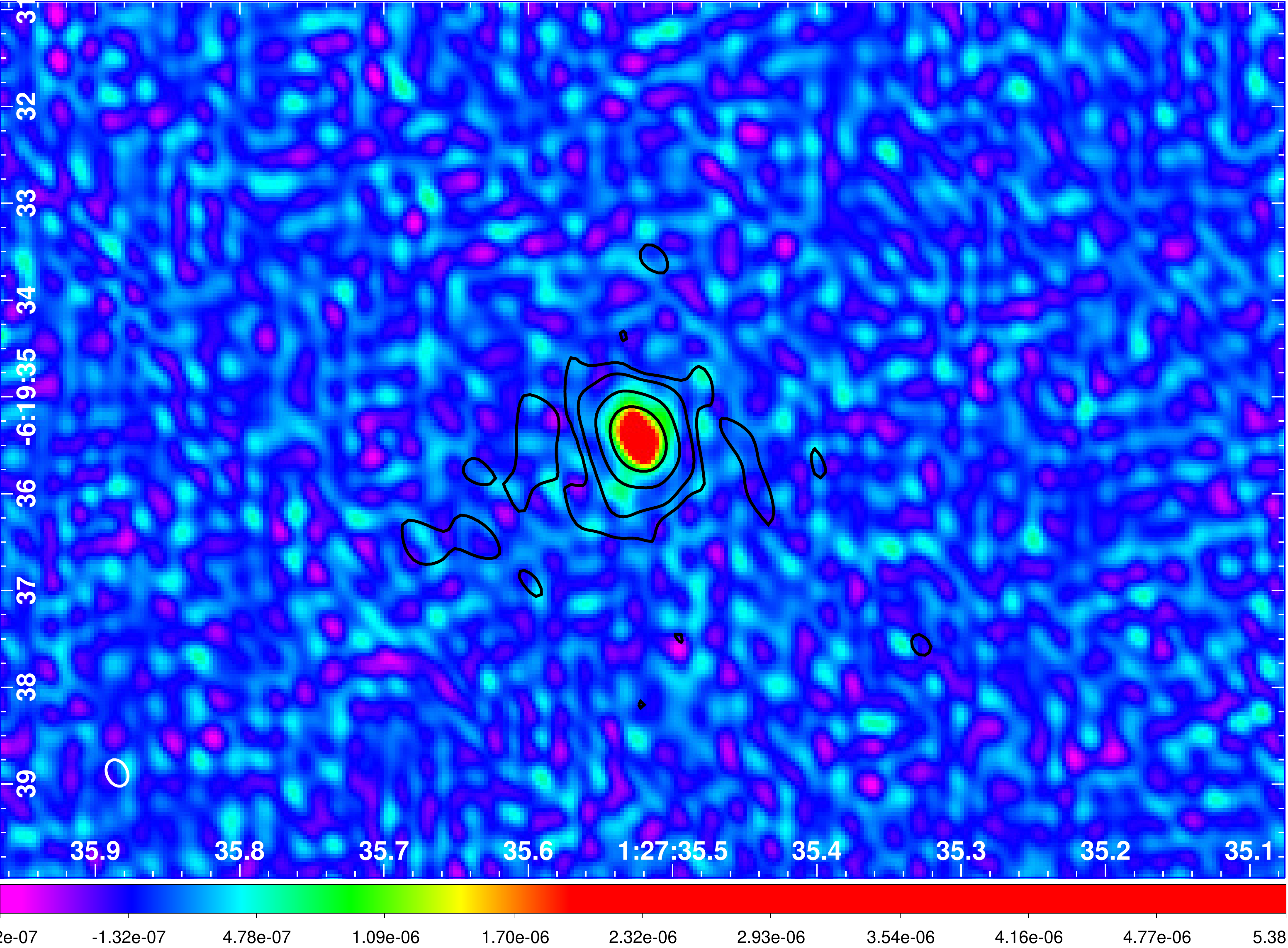}}\hfill
\subfloat[SBS 0940+544 VLA 5 GHz image with contour levels of $(\sqrt{2})^n$ times the RMS noise, where $n=4$, 5, and 6. \label{fig:radioi}]{\includegraphics[width=0.48\textwidth]{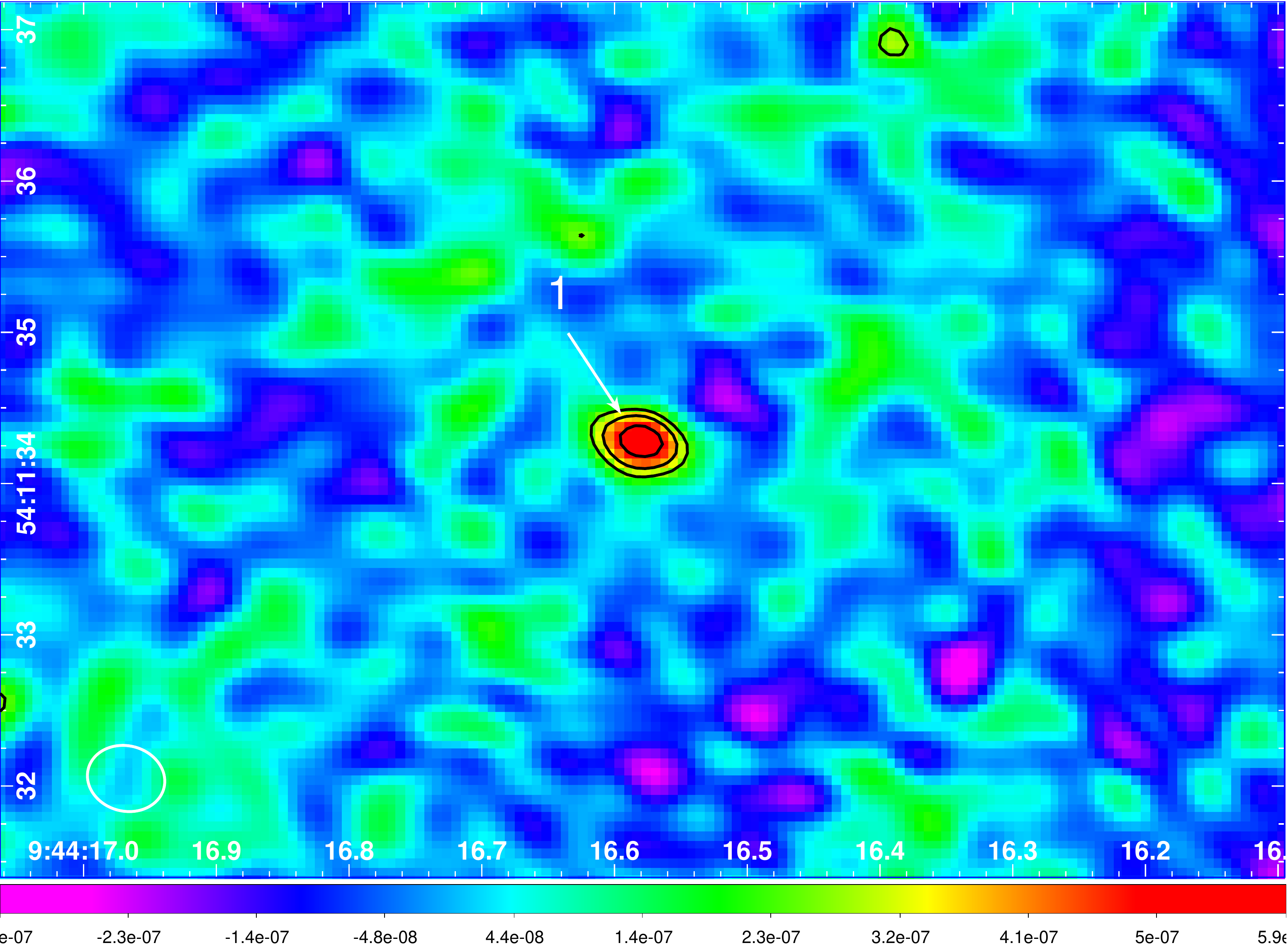}}\hfill
\subfloat[SBS 0940+544 VLA 7.4 GHz image with the 5 GHz contour levels overlaid. \label{fig:radioj}]{\includegraphics[width=0.48\textwidth]{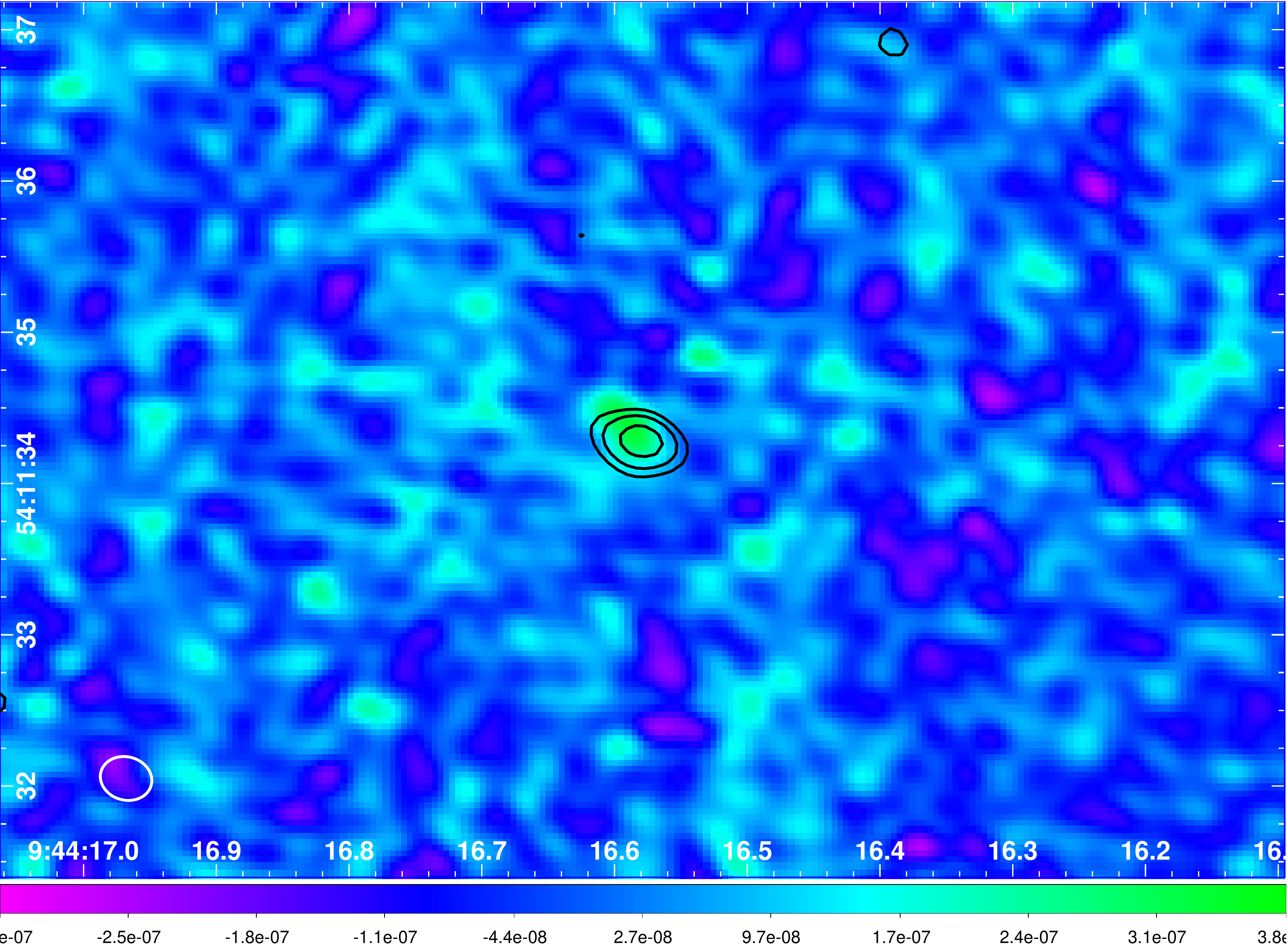}}
\caption{VLA 5 GHz (left) and 7.4 GHz (right) images of the galaxies in our sample. The beam sizes are shown in the lower left corners.}
\label{fig:radio}
\end{figure*}
\begin{deluxetable*}{cccrc}
\tabletypesize{\footnotesize}
\tablecaption{Imaging Parameters for VLA Observations}
\tablewidth{0pt}
\tablehead{
\colhead{Galaxy}  & \colhead{$\nu$}  & \colhead{Synthesized Beam} & \colhead{P.A.} & \colhead{rms Noise} \\
\colhead{ }  & \colhead{(GHz)}  & \colhead{(arcsec)} & \colhead{(deg)} & \colhead{($\mu$Jy beam$^{-1}$)} \\
\colhead{(1)}  & \colhead{(2)}  & \colhead{(3)} & \colhead{(4)} & \colhead{(5)}   }
\startdata
II Zw 70              & 5.0 & 0.52 $\times$ 0.43  & $-78.7$ &  4.5  \\
\nodata             & 7.4 & 0.36 $\times$ 0.28  & $-83.0$ &    4.3\\
Haro 3                & 5.0 & 0.47 $\times$ 0.41  & $30.2 $ &   10.4\\
\nodata             & 7.4 & 0.32 $\times$ 0.27  & $40.5  $ &   5.7\\
Haro 9                & 5.0 & 0.51 $\times$ 0.43  & $-65.1$ &   5.6\\
\nodata             & 7.4 & 0.34 $\times$ 0.28  & $-74.2$ &   4.4\\
Mrk 996             & 5.0 & 0.60 $\times$ 0.39  & $27.9 $ &   4.6\\
\nodata             & 7.4 & 0.29 $\times$ 0.21  & $25.2  $ &   7.6\\
SBS 0940+544        & 5.0 & 0.52 $\times$ 0.43  & $75.9 $ &   5.9\\
\nodata             & 7.4 & 0.35 $\times$ 0.29  & $75.1  $ &   5.2
\enddata
\tablecomments{Column 1: galaxy name. 
Column 2: rest frequency.  
Column 3: beam major axis $\times$ beam minor axis.
Column 4: beam position angle.  
Column 5: RMS noise of the image. 
All images were produced using natural weighting for maximum sensitivity.}
\label{tab:vlaim}
\end{deluxetable*}

\subsection{Optical Images} \label{sec:optobs}

We obtained optical images of our target galaxies for comparison with the X-ray and radio observations. SDSS color composite images of the BCDs are shown in Figure \ref{fig:sdss}.  We also retrieved archival {\it Hubble Space Telescope (HST)} WFPC2 broadband images of three galaxies (Haro 3, Haro 9, and Mrk 996) from the Hubble Legacy Archive.  The absolute astrometry of the {\it HST} images was adjusted to match the SDSS using common sources in each image.  The astrometric corrections were $\lesssim 0\farcs2$.

\section{Analysis and Results} \label{sec:results}

\subsection{Host Galaxies and Star Formation Rates} \label{sec:hostgals}

The five galaxies in our sample (see Figure \ref{fig:sdss}) span a stellar mass range of ${\sim}10^{7} - 10^{8.4}$ \msun, with a median of ${\sim}10^8$ \msun.  It is worth noting that these galaxies have more irregular morphologies and significantly lower stellar masses than previous samples of dwarf galaxies found to host AGNs \citep[e.g.,][]{reines13}.  The galaxies are also physically small, with half-light radii less than 1 kpc.  All of the galaxies are relatively nearby, with distances in the range of ${\sim}17 \textrm{-} 27$ Mpc (using the \texttt{zdist} parameter in the NSA and adopting $h_0=0.73$). See Table \ref{tab:sample} for information on the individual galaxies.

In general, BCDs are experiencing regions of intense star formation.  We estimate the star formation rates (SFRs) of our target BCD galaxies using far-UV (FUV; 1528 \AA) and mid-infrared (IR; 25 $\mu$m) luminosities via
\begin{equation} \label{eq:sfrs}
\begin{gathered}
 \textrm{log SFR}(\textrm{\msun~yr}^{-1}) = \textrm{log }L\textrm{(FUV)}_{\textrm{corr}} - 43.35 \\
 \textrm{log }L \textrm{(FUV)}_{\textrm{corr}} = L\textrm{(FUV)}_{\textrm{obs}} + 3.89 L\textrm{(25 $\mu$m)}
\end{gathered}
\end{equation}
\citep{kennicutt12, hao11, murphy11}. We obtain FUV magnitudes from the \textit{Galaxy Evolution Explorer} (\textit{GALEX}), and 22 $\mu$m magnitudes from the \textit{Wide-field Infrared Survey Explorer} (\textit{WISE}) \citep{wright10}. While the \citet{hao11} relation uses 25 $\mu$m luminosities from the \textit{Infrared Astronomical Satellite} (\textit{IRAS}), only three of the galaxies in our sample had \textit{IRAS} detections while all five were detected by \textit{WISE}. Therefore, we use 22 $\mu$m flux densities as a proxy for 25 $\mu$m flux densities as this ratio is expected to be of order unity \citep{jarrett13}. The resulting estimates for the SFRs are summarized in Table \ref{tab:sfrs}.
The SFRs span ${\sim} 0.03 \textrm{-} 0.87$ \msun~yr$^{-1}$, with a median of ${\sim} 0.27$ \msun~yr$^{-1}$. The uncertainty in this method is ${\sim} 0.13$ dex.

The specific star formation rates (sSFR=SFR/$M_\star$) of these BCDs are quite high.  Using SFRs calculated above and stellar masses from the NSA, we estimate log sSFRs in the range $-8.6$ to $-7.9$ yr$^{-1}$. This is comparable to or greater than other dwarf galaxies with AGN candidates such as Mrk 709 and Henize 2-10 \citep[which have log sSFRs of $-8.6$ and $-9.3$ yr$^{-1}$, respectively;][]{reines14,reines11} and orders of magnitude higher than that of the Large Magellanic Cloud \citep[log~sSFR $\sim {-}10$~yr$^{-1}$;][]{whitney08,vandermarel02}.

\begin{deluxetable*}{cccccccc}
\tabletypesize{\footnotesize}
\tablecaption{Host Galaxy SFRs and Expected Luminosity from XRBs}
\tablewidth{0pt}
\tablehead{
\colhead{Galaxy} & \colhead{FUV} & \colhead{log L(FUV)}  & \colhead{W$_{22}$} & \colhead{log L(22 $\mu$m)} & \colhead{SFR$_{\rm FUV}$} & \colhead{log sSFR} & \colhead{log L$^{\rm XRB}_{\rm 2-10~keV}$}  \\
\colhead{ } & \colhead{(mag)} & \colhead{(erg s$^{-1}$)} & \colhead{(mag)} & \colhead{(erg s$^{-1}$)} & \colhead{(\msun~yr$^{-1}$)} & \colhead{} & \colhead{(erg s$^{-1}$)} \\
\colhead{(1)} & \colhead{(2)} & \colhead{(3)} & \colhead{(4)} & \colhead{(5)} & \colhead{(6)} & \colhead{(7)} & \colhead{(8)}}
\startdata
II Zw 70  		 &  14.79  & 42.6 & 4.92  & 41.7 & 0.27 & $-7.9$ & 39.0  \\
Haro 3    		 &  14.30  & 42.7 & 2.50  & 42.6 & 0.87 & $-8.1$ & 39.4  \\
Haro 9    		 &  13.53  & 43.0 & 3.96  & 42.1 & 0.69 & $-8.5$ & 39.4  \\
Mrk 996 		 &  16.47  & 42.0 & 4.85  & 41.8 & 0.15 & $-8.6$ & 38.8  \\
SBS 0940+544     &  17.79  & 41.7 & 8.07  & 40.8 & 0.03 & $-8.5$ & 38.2
\enddata
\tablecomments{Column 1: galaxy name.  
Column 2: FUV AB magnitudes from \textit{GALEX} (through the NSA).
Column 3: log FUV luminosities.
Column 4: \textit{WISE} magnitudes\footnote{Similarly to the NSA, SBS 0940+544 appears in \textit{WISE} as two sources; the bright central region and the trailing tail region. The values reported here are for both sources, combined.}. 
Column 5: log 22 $\mu$m luminosities.
Column 6: estimated SFRs from {\it GALEX} and {\it WISE} data.
Column 7: log specific SFRs (sSFR=SFR/$M_\star$) in units of yr$^{-1}$.
Column 8: log total expected 2-10 keV luminosity from high-mass XRBs.
}
\label{tab:sfrs}
\end{deluxetable*}

\subsection{Hard X-ray Sources} \label{sec:xrayemis}

We search for hard X-ray sources to identify the presence of an accreting BH, compared to star formation which produces high-energy radiation predominantly in the soft X-ray band. To identify hard X-ray sources in our sample of BCDs, we re-run \textit{\tt wavdetect} on (astrometrically corrected) hard images of the S3 chip filtered from 2-7~keV. We adopt wavelet scales of 1.0, 1.4, 2.0, 2.8, 4.0 pixels, a point spread function map describing the 39\% enclosed energy fraction at 4~keV, and a significance threshold of 10$^{-6}$ (corresponding to approximately one expected false point source detection across the S3 chip).  We then restrict the list of hard X-ray sources identified by {\tt wavdetect} to those that are located within 3$r_{50}$ (Petrosian 50\% light radii) of the galaxy optical center.  How we assess the statistical significance of each hard X-ray source detection is described below.

We extract source counts within circular apertures centered on each source, with aperture radii of 3 pixels ($1\farcs5$).  These apertures correspond to the 90\% enclosed energy fraction at 4.5~keV at the S3 aimpoint (all X-ray sources are located $<0\farcm5$ from the aimpoint).  We adopt these (relatively small) 90\% circular apertures to avoid contamination from nearby sources and we visually examine each image to confirm that there is no contamination within any source aperture.  

The number of background counts per pixel is estimated by using a circular aperture with  25 arcsec radius centered on a manually selected source-free region of the chip near each galaxy.   We consider a source to be detected if the source counts are above the background level within each source aperture at the $>$95\% confidence level, following the \citet{kraft91} Bayesian formalism for Poisson counting statistics in the presence of a background.  Only one {\tt wavdetect} source in  Haro 3 fails to meet this detection criterion.  

Finally, after subtracting the expected number of background counts in each source aperture from the total counts, a 90\% aperture correction is applied to calculate the net counts and net count rates from each source.  A total of 10 hard X-ray sources are detected in four galaxies (none are found in II Zw 70), and their properties are reported in Table \ref{tab:xray}. Note that we exclude an additional source (X2 in SBS~0940+544) from further analysis, as it is likely a background quasar. This source is ${\sim}20$ arcsec away from the main body of the galaxy and it appears in the Milliquas catalog of \cite{flesch15} under object name SDSS~J094418.57+541141.4 with a photometric redshift of ${\sim}1.6$.

The uncertainty of each hard X-ray position is quoted as the radius of the 95\% error circle, as determined from Equation 5 of \citet{hong05}, who ran simulations to determine the positional accuracy of sources detected by  {\tt wavdetect} as a function of counts and location on the ACIS detector.   Error bars on counts are quoted at the 90\% confidence level.  For sources with $<$10 total counts, errors are calculated following \citet{kraft91}, which incorporates the number of background counts in each source aperture.  If  $\geq$10 total counts, then we assume the background is negligible (all source apertures contain $<$0.5 background counts), and we adopt the 90\% confidence interval from \citet{gehrels86}.

Unabsorbed hard X-ray fluxes are calculated from 2-10~keV using  the Portable, Interactive Multi-Mission Simulator (PIMMS)\footnote{\url{http://heasarc.gsfc.nasa.gov/docs/software/tools/pimms.html}}.
We adopt the Galactic column density from the \citet{dickey90} maps, and we assume a power-law spectral model with photon index $\Gamma=1.8$, which is typical for low-luminosity AGN \citep{ho08,ho09} and ultraluminous X-ray sources \citep{swartz08}. Unabsorbed fluxes and corresponding luminosities are reported in Table \ref{tab:xray}.  We ignore any potential absorption that is intrinsic to the source and/or host dwarf galaxy, hence these fluxes and luminosities should be considered lower limits.  

Finally, we determine that there is a low probability of any of the detected hard X-ray sources in Table \ref{tab:xray} being a chance alignment of a resolved X-ray source from the cosmic X-ray background. We estimate the minimum 2-10~keV (unabsorbed) flux sensitivity of each observation that corresponds to detecting 3 hard X-ray counts (i.e., the 1-sided 95\% confidence limit assuming no background), given the exposure time of each observation and the Galactic column density toward each galaxy (and assuming $\Gamma=1.8$).  The flux sensitivities range from $S_{\rm min}$=1-4$\times10^{-15}$ erg s$^{-1}$ cm$^{-2}$ for our five \textit{Chandra} observations.    We then use the resolved cosmic X-ray background $\log N - \log S$ relation of \citet{moretti03} to estimate $N\left(>S_{\rm min}\right)$, the cumulative number of 2-10~keV X-ray sources per deg$^2$ with a flux $>S_{\rm min}$.  Based on these $N\left(>S_{\rm min} \right)$ estimates, the expected number of background sources to fall within $3r_{50}$ of each galaxy's optical center is small, ranging from 0.01-0.2 for each of the five galaxy targets.

Our observed 2-10 keV X-ray luminosities range from log $L_{\rm 2-10 keV} ({\rm erg~s}^{-1}) \sim 38 - 39.5$ (Table \ref{tab:xray}). Most of our X-ray sources lack optical counterparts, with the exceptions of X1 in Haro 9 and X2 in SBS~0940+544. Two of the galaxies in our sample (with archival {\it Chandra} data) have previously detected X-ray sources. \cite{georgakakis11} searched Mrk~996 for indications of an intermediate mass BH using {\it Chandra} data, finding no conclusive evidence for one. They detected the same X-ray source presented here (X1) with $L_{\rm 0.3-10 keV} = 1.2 \times 10^{39}$ erg s$^{-1}$, about an order of magnitude higher than the 2-10 keV luminosity we report, as well as another less luminous off-nuclear X-ray source (that we did not detect in the hard 2-10 keV band) with $L_{\rm 0.3-10 keV} = 1.8 \times 10^{38}$ erg s$^{-1}$. \cite{prestwich13} detected an X-ray source in SBS~0940+544 (X1 in this paper) with $L_{\rm 0.3-8 keV} = 1.26 \times 10^{39}$ erg s$^{-1}$. While we use the same {\it Chandra} data as \cite{georgakakis11} and \cite{prestwich13}, the reported source luminosities differ due to the different energy ranges considered.

\begin{deluxetable*}{crrccccc}
\tabletypesize{\footnotesize}
\tablecaption{Hard X-ray Sources}
\tablewidth{0pt}
\tablehead{
\colhead{Source ID}  & \colhead{R.A.} & \colhead{Decl.}  & \colhead{$p_{err}$} & \colhead{Net Counts} &
\colhead{$N_{\rm bg}$ Counts} & \colhead{$F_{\rm 2-10~keV}$}  & \colhead{log $L_{\rm 2-10~keV}$} \\
\colhead{ } & \colhead{(deg)} & \colhead{(deg)} & \colhead{(arcsec)} & \colhead{ } & \colhead{ } & \colhead{($10^{-15}$ erg s$^{-1}$ cm$^{-2}$)} & \colhead{(erg s$^{-1}$)} \\
\colhead{(1)} & \colhead{(2)} & \colhead{(3)} & \colhead{(4)} & \colhead{(5)} & \colhead{(6)} & \colhead{(7)} & \colhead{(8)}  }
\startdata
Haro 3$-$X1                & 161.346025 &    55.954587 &  0.34 & $76.39 \pm 14.94$           &   0.12 &    90.07  &        39.5  \\
Haro 3$-$X2                & 161.341482 & 55.961223     & 0.42 & $13.11^{+7.69}_{-5.34}$ &   0.12 &    15.46  &        38.7  \\
Haro 3$-$X3                & 161.341573 & 55.959588     & 0.47 & $8.65    \pm 4.75$             &   0.12 &    10.20   &       38.5  \\
Haro 3$-$X4                & 161.346679 & 55.961958     & 0.49 & $7.64    \pm 4.45$             &   0.12 &     9.01    &        38.5  \\
Haro 3$-$X5                & 161.354910 & 55.961723     & 0.94 & $4.31    \pm 3.46$             &   0.12 &     5.08    &        38.2  \\
Haro 9$-$X1                & 191.322023 & 27.125507     & 0.35 & $45.49 \pm 11.68$           &   0.08 &     56.59  &        39.4  \\
Haro 9$-$X2                & 191.320506 & 27.126089     & 0.39 & $20.19 \pm 8.02$              &   0.08 &     25.12  &       39.0  \\
Haro 9$-$X3 		   & 191.320973 & 27.126501     & 0.39 & $19.28 \pm 7.86$              &   0.08 &     23.98  &        39.0  \\
Mrk 996$-$X1             & 21.898420    & $-6.325036$ & 0.52 & $6.72   \pm 4.51$              &   0.47 &     2.66     &       38.1  \\
SBS 0940+544$-$X1 & 146.068446  & 54.192835    & 0.40 & $16.58 \pm 7.34$              &   0.12 &     20.69  &        39.2
\enddata
\tablecomments{Column 1: hard X-ray source identification.  
Column 2: right ascension.  
Column 3: declination.  
Column 4: 95\% positional uncertainty.  
Column 5: net counts in the 2-7 keV energy range, after applying a 90\% aperture correction.  Error bars represent 90\% confidence intervals.  
Column 6: number of background counts expected within each source extraction circle (after applying a 90\% aperture correction). 
Column 7: 2-10 keV flux corrected for Galactic absorption. 
Column 8: log 2-10 keV luminosity corrected for Galactic absorption.
}
\label{tab:xray}
\end{deluxetable*}

\subsubsection{Expected Contribution from X-ray Binaries} \label{sec:xrbs}

The luminosities of our detected X-ray sources ($L_{\rm 2-10 keV} \sim 10^{38-39.5}$ erg s$^{-1}$) are consistent with either massive BHs accreting well below their Eddington luminosity (e.g., a $\sim 10^4 M_\odot$ BH with an Eddington ratio of $\sim 10^{-3}$), or stellar-mass BHs (or neutron stars) in XRBs radiating at a significant fraction of their Eddington luminosity. X-ray emission from high-mass XRBs is known to increase as a function of SFR for late-type galaxies \citep{grimm02,grimm03,gilfanovetal04,mineo12}, whereas low-mass XRBs dominate the XRB population in early-type galaxies and scale with stellar mass \citep{gilfanov04,humphreybuote08,lehmer10,lehmer14}. As our target galaxies are generally late-type (irregular) and have high SFRs relative to their stellar mass (see Section \ref{sec:hostgals}), most of the expected X-ray emission will likely be due to high-mass XRBs.

Enhanced X-ray emission relative the expected contribution from XRBs {\it could} indicate the presence of a massive BH, however this is neither a necessary nor sufficient condition. For example, a highly sub-Eddington massive BH would likely not contribute significantly to the cumulative X-ray emission, and enhanced X-ray emission can originate from very luminous stellar-mass XRBs \citep{brorby14}.

Nevertheless, we estimate the expected cumulative 2-10 keV luminosity from XRBs in each galaxy using the relation in \cite{lehmer10} which accounts for high-mass XRBs:

\begin{equation} \label{eq:expxrb}
L_{\textrm{HX}} =
	\begin{cases}
		10^{(39.57 \pm 0.11)} \textrm{SFR}^{(0.94 \pm 0.15)} & \textrm{SFR} \lesssim 0.4 \textrm{\msun~yr$^{-1}$} \\
		10^{(39.49 \pm 0.21)} \textrm{SFR}^{(0.74 \pm 0.12)} & \textrm{SFR} \gtrsim 0.4 \textrm{\msun~yr$^{-1}$.} \\
	\end{cases}
\end{equation}

\noindent
We use the SFRs derived in Section \ref{sec:hostgals} (Table \ref{tab:sfrs}). We note that the uncertainty in the SFRs is $\sim 0.13$ dex \citep{hao11} and the 1$\sigma$ scatter in the \cite{lehmer10} relationship is $\sim 0.4$ dex. We find the expected 2-10 keV luminosities from XRB emission to be in the range  $\sim 10^{38.2 - 39.4}$ erg s$^{-1}$ (see Table \ref{tab:sfrs}).

\begin{figure}[h!]
\centering
\includegraphics[width=0.48\textwidth]{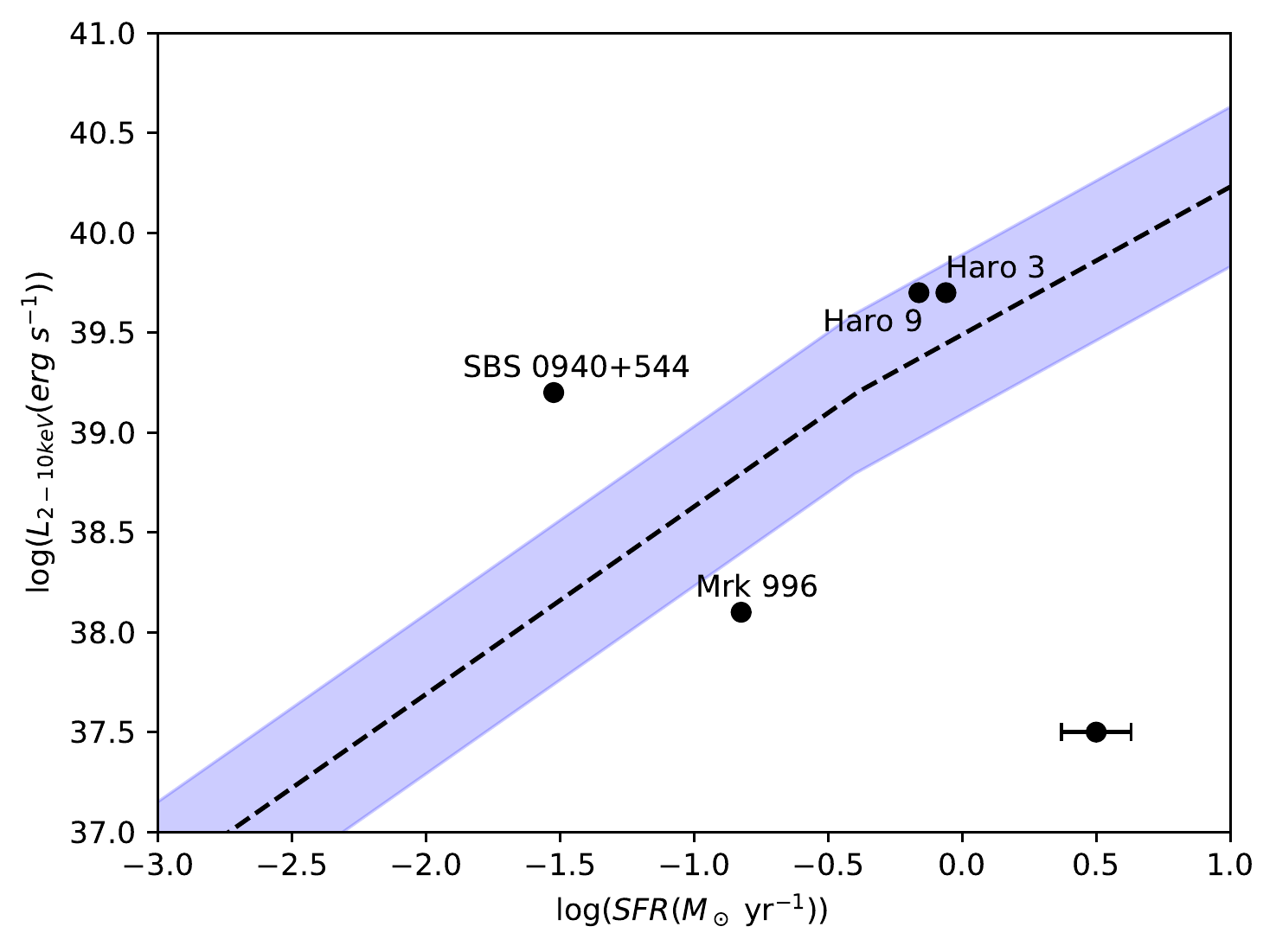}
\caption{X-ray luminosity vs.\ SFR for the galaxies. The points represent the observed X-ray luminosities (cumulative from the point sources), while the black dashed line and shaded blue area represent the expected X-ray luminosity and the 1$\sigma$ scatter using the relation from \cite{lehmer10}. The uncertainty of ${\sim} 0.13$ dex in the SFR is shown in black in the lower right-hand corner. Note that II Zw 70 is absent from this plot as we did not significantly detect any hard X-ray sources.}
\label{fig:expxrbalt}
\end{figure}

The observed X-ray luminosities for the galaxies (cumulative from point sources) are generally within the 1$\sigma$ scatter of the \cite{lehmer10} relationship, with the exception of SBS 0940+544 which has an X-ray luminosity ${\sim} 3\sigma$ higher than would be expected from star formation (see Figure \ref{fig:expxrbalt}; note that II Zw 70 is not plotted as it did not have any observed hard X-ray sources). This discrepancy is unlikely to be due to metallicity, as the metallicities of SBS 0940+544 and Mrk 996 are comparable (at most ${\sim}0.2$ dex difference, from the NSA values) and Mrk 996 has slightly \textit{lower} X-ray luminosity than expected. However, a combination of stochastic effects (regarding XRB formation in galaxies) and the relatively low SFR of SBS~0940+544 could explain this unexpectedly high X-ray luminosity.  

If the X-ray source in SBS~0940+544 were a massive BH, we would expect corresponding radio emission as predicted using the fundamental plane of BH activity from \cite{merloni03}. The luminosity of X1 in SBS~0940+544 is $L_{\rm 2-10~keV} \sim 10^{39.2}$ erg s$^{-1}$ (see Table \ref{tab:xray}). Assuming a BH mass of log~$(M_{\rm BH}/M_\odot) \sim 3.3 \pm 0.6$ \citep{reines15}, we would expect a radio luminosity of $L_{\rm 5GHz} \sim 10^{33.4}$ erg s$^{-1}$ where the \citet{merloni03} relation has a scatter of $\sim$ 0.9 dex. Given that the expected radio emission is within our detection limits, and we do not detect radio emission from this source, we do not consider a massive BH to be a likely explanation.

\subsection{Compact Radio Sources} \label{sec:rademis}

\begin{deluxetable*}{crrrrcccc}
\tabletypesize{\footnotesize}
\tablecaption{Compact Radio Sources}
\tablewidth{0pt}
\tablehead{
\colhead{Source ID}  & \colhead{R.A.} & \colhead{Decl.}  & \colhead{P$_{\rm 5GHz}$} & \colhead{P$_{\rm 7.4GHz}$}& \colhead{$F_{\rm 5GHz}$} & \colhead{$F_{\rm 7.4GHz}$} & \colhead{log $L_{\rm 5GHz}$} & \colhead{log $L_{\rm 7.4GHz}$}  \\
\colhead{ } & \colhead{(deg)} & \colhead{(deg)} & \colhead{($\mu$Jy/beam)} & \colhead{($\mu$Jy/beam)} & \colhead{($\mu$Jy)} & \colhead{($\mu$Jy)} & \colhead{ } & \colhead{ } \\
\colhead{(1)} & \colhead{(2)} & \colhead{(3)} & \colhead{(4)} & \colhead{(5)} & \colhead{(6)} & \colhead{(7)} & \colhead{(8)}& \colhead{(9)}  }
\startdata
II Zw 70$-$R1		  & 222.735642 & 35.572258  & 181  & 91    & $360     \pm 32   $ & $331     \pm 33    $ & 35.9 & 36.0 \\
Haro 3$-$R1   		  & 161.344137 & 55.960794  & 119  & 89    & $93      \pm 5    $ & $90      \pm 5     $ & 35.2 & 35.4 \\
Haro 3$-$R2   		  & 161.343988 & 55.961128  & 80   & 46    & $77      \pm 8    $ & $101     \pm 12    $ & 35.1 & 35.4 \\
Haro 3$-$R3a 		  & 161.341874 & 55.961178  & 725  & 526   & $1500    \pm 83   $ & $1540    \pm 86    $ & 36.4 & 36.6 \\ 
Haro 3$-$R3b 	  & 161.341309 & 55.960978  & 664  & 432   & $2080    \pm 112  $ & $2080    \pm 113   $ & 36.5 & 36.7 \\ 
Haro 9$-$R1   		  & 191.321930 & 27.125492  & 55   & 38    & $24      \pm 3    $ & $38      \pm 3     $ & 34.7 & 35.1\\
Haro 9$-$R2   		  & 191.321275 & 27.124692  & 44   & 37    & $45      \pm 5    $ & $52      \pm 10    $ & 34.9 & 35.2 \\
Haro 9$-$R3   		  & 191.320058 & 27.125258  & 53   & 32    & $39      \pm 5    $ & $26      \pm 4     $ & 34.9 & 34.9 \\
Mrk 996$-$R1		  & 21.898009 & $-6.326506$ & 359  & 230   & $590     \pm 31   $ & $549     \pm 31    $ & 36.2 & 36.3 \\
SBS 0940+544$-$R1     & 146.069071 & 54.192850  & 42   & 24    & $23      \pm 2    $ & $19      \pm 2     $ & 35.0 & 35.1 
\enddata
\tablecomments{Column 1: radio source identification.  
Column 2: right ascension.  
Column 3: declination. 
Columns 4-5: peak flux values of each source at 5 GHz and 7.4 GHz. 
Columns 6-7: flux densities ($F_\nu$) at 5 GHz and 7.4 GHz. 
Columns 8-9: log luminosities ($\nu L_\nu$) at 5 GHz and 7.4 GHz in units of erg s$^{-1}$.
}
\label{tab:radio}
\end{deluxetable*}

Compact radio emission is detected in all five of our target galaxies (see Figure \ref{fig:radio}).
We use the \texttt{detect\_sources} and \texttt{deblend\_sources} functions from the Astropy-affiliated Photutils Python package to select our sources. We use multiples of $\sigma$, where $\sigma$ is the RMS noise (see Table \ref{tab:vlaim}) as thresholds for source detection, adapting the threshold to each image as so to eliminate spurious detections. We then relax these thresholds to measure the flux densities to ensure we capture all the emission from the source. 

While the flux densities have an intrinsic ${\sim} 5\%$ error due to calibration\footnote{\url{https://science.nrao.edu/facilities/vla/docs/manuals/oss/performance/fdscale}}, we also introduce uncertainty into our measurements via the exact choice of background we use for each source. To account for this, we take take multiple measurements of each source, varying said background. We report the average flux densities from these measurements, and include both the intrinsic ${\sim} 5\%$ error and the standard deviation of these measurements in our reported uncertainties. 

We identify a total of 10 compact radio sources across the five galaxies, with luminosities in the range $4.8 \times 10^{34} - 5.2 \times 10^{36}$ erg~s$^{-1}$. Due to uncertainties in flux densities and limited frequency coverage, we are not able to reliably determine the spectral indices for most of our sample.
Two of the galaxies in our sample have previously detected compact radio sources. \cite{cox01} combined 1.4 GHz radio and optical observations to analyze the polar ring of II Zw 71, the companion galaxy to II Zw 70; they also detected the radio source we denote as R1 in this work. \cite{johnson04} used 8.3 GHz and 23 GHz radio observations to study the ongoing star formation in Haro~3 and detected the same radio sources presented here.

\subsubsection{Comparison to Thermal H{\footnotesize II} Regions and SNRs} \label{sec:radcomp}

While we are primarily interested in detecting potential radio emission from accreting massive BHs, we must also consider alternative origins for the observed compact radio emission since we are dealing with lower luminosities compared to more massive systems.  We are mainly concerned about free-free emission from dense H{\footnotesize II} regions and synchrotron emission from supernova remnants.

We first consider the possibility that the radio emission is entirely thermal and emanating from dense H{\footnotesize II} regions associated with extragalactic massive star clusters.  Under this assumption, we estimate the ionizing flux, $Q_{Lyc}$, of the radio sources following \cite{condon92}:  
\begin{multline} \label{eq:qlyc}
\left(\frac{Q_{Lyc}}{\textrm{s}^{-1}}\right) \gtrsim 6.3 \times 10^{52} \left(\frac{T_e}{10^4 \textrm{ K}}\right)^{-0.45} \left(\frac{\nu}{\textrm{GHz}}\right)^{0.1} \\ \times  \left(\frac{L_{\textrm{$\nu$, thermal}}}{10^{27} \textrm{ erg s}^{-1} \textrm{ Hz}^{-1}}\right).
\end{multline}

\noindent
We assume an electron temperature $T_e \sim 10^4$ K for all sources and use the 7.4 GHz luminosities from Table \ref{tab:radio}.
Adopting $Q_{\rm Lyc} = 10^{49}$ s$^{-1}$ as the ionizing flux of a typical O-type main sequence star (O7.5 V star; \citealt{vacca96}), we find ionizing fluxes corresponding to the equivalent of $\sim$ 80 to 5410 O-type stars (median value of 250 O-type stars). The calculated ionizing fluxes are consistent with thermal radio sources associated with young massive star clusters in other star-forming dwarf galaxies \citep[e.g.,][]{johnson04,reines08,aversa11,kepley14}.

We also calculate the expected galaxy-wide ionizing fluxes using the UV-derived SFRs in Section \ref{sec:hostgals} and Equation 2 in \cite{kennicutt98}, and compare these values to those from adding up the individual radio sources.  We find the two to generally be in agreement, differing by at most a factor of ${\sim}$2 with the exception of Haro 9. The UV-derived galaxy-wide ionizing flux of Haro 9 is greater by a factor of ${\sim}$18 compared to the sum of the radio sources.  This discrepancy is likely due to the large amount of extended emission in Haro 9 (see Figures \ref{fig:radioe}, \ref{fig:radiof}), which is included in the galaxy-wide UV-derived ionizing flux but excluded when simply adding up the ionizing fluxes of the individual sources.

Next we compare the luminosities of our detected radio sources with known radio SNRs.  We assume a typical SNR spectral index of $\alpha = 0.7$, where $F_{\nu} \propto \nu^{-\alpha}$, and convert our 7.4 GHz flux densities (Table \ref{tab:radio}) to 1.45 GHz spectral luminosities to compare with the SNR spectral luminosities in \cite{chomiuk09}.

The corresponding 1.45 GHz spectral luminosities span a range of ${\sim} 2 \times (10^{25} \textrm{-} 10^{27})$ erg s$^{-1}$ Hz$^{-1}$, which are consistent with or slightly above the radio SNRs in \citet{chomiuk09} ($L_{\rm 1.4 GHz} {\sim} 10^{23}$ - $10^{27}$ erg s$^{-1}$ Hz$^{-1}$).  We therefore conclude that the compact radio sources in our sample are not so luminous as to exclude star-formation-related emission in the form of H{\footnotesize II} regions and/or SNRs.

\subsection{A Candidate Low-luminosity AGN in Haro~9} \label{sec:candidate}

\begin{figure*}
\subfloat[II Zw 70 optical image from the SDSS (\textit{g} band). \label{fig:opticala}]{\includegraphics[width=0.48\textwidth]{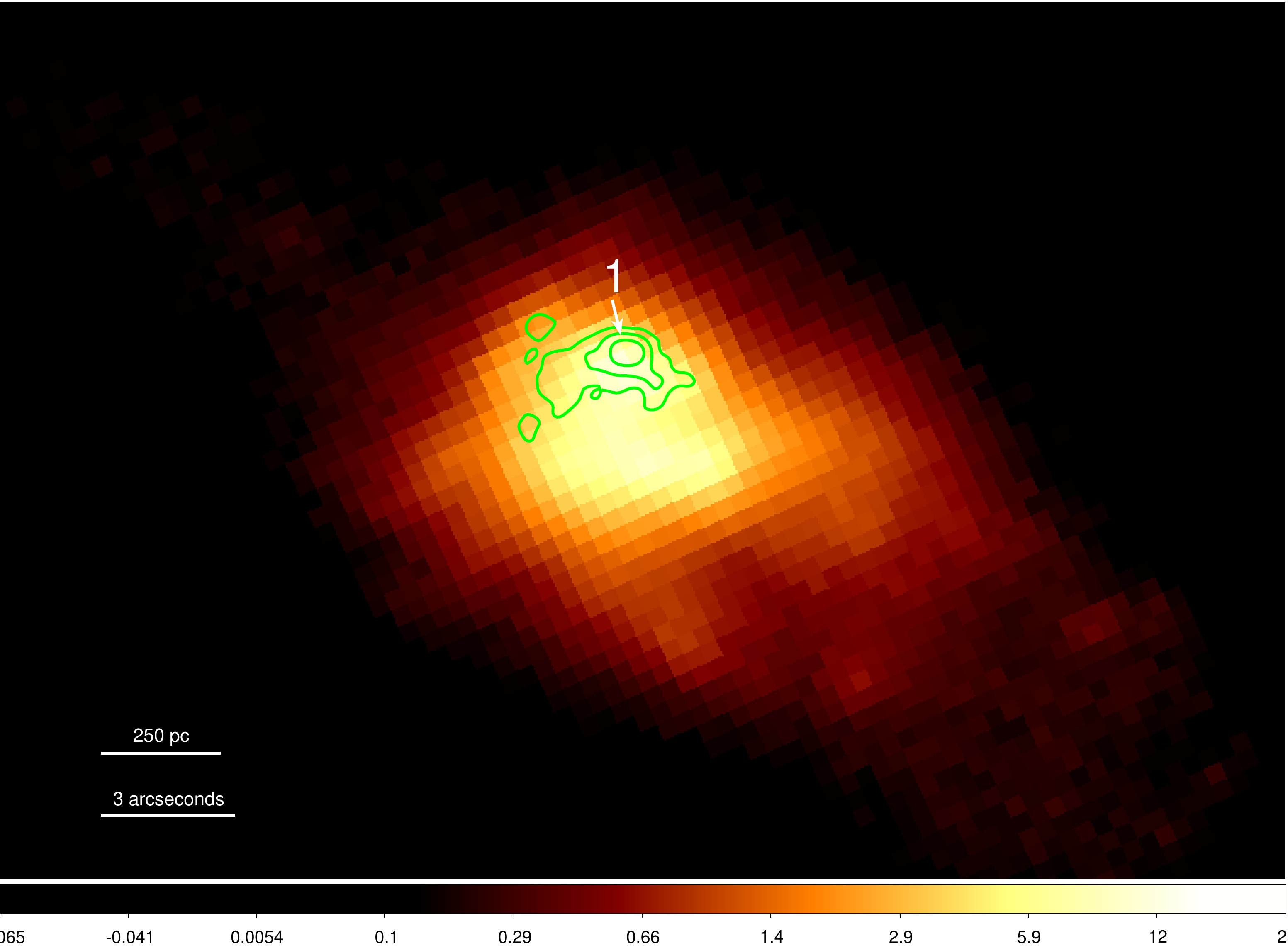}}\hfill
\subfloat[Haro 3 optical image from {\it HST}/WFPC2 (F606W filter). Note that X-ray sources X1 and X5 are out of view.  \label{fig:opticalb}]{\includegraphics[width=0.48\textwidth]{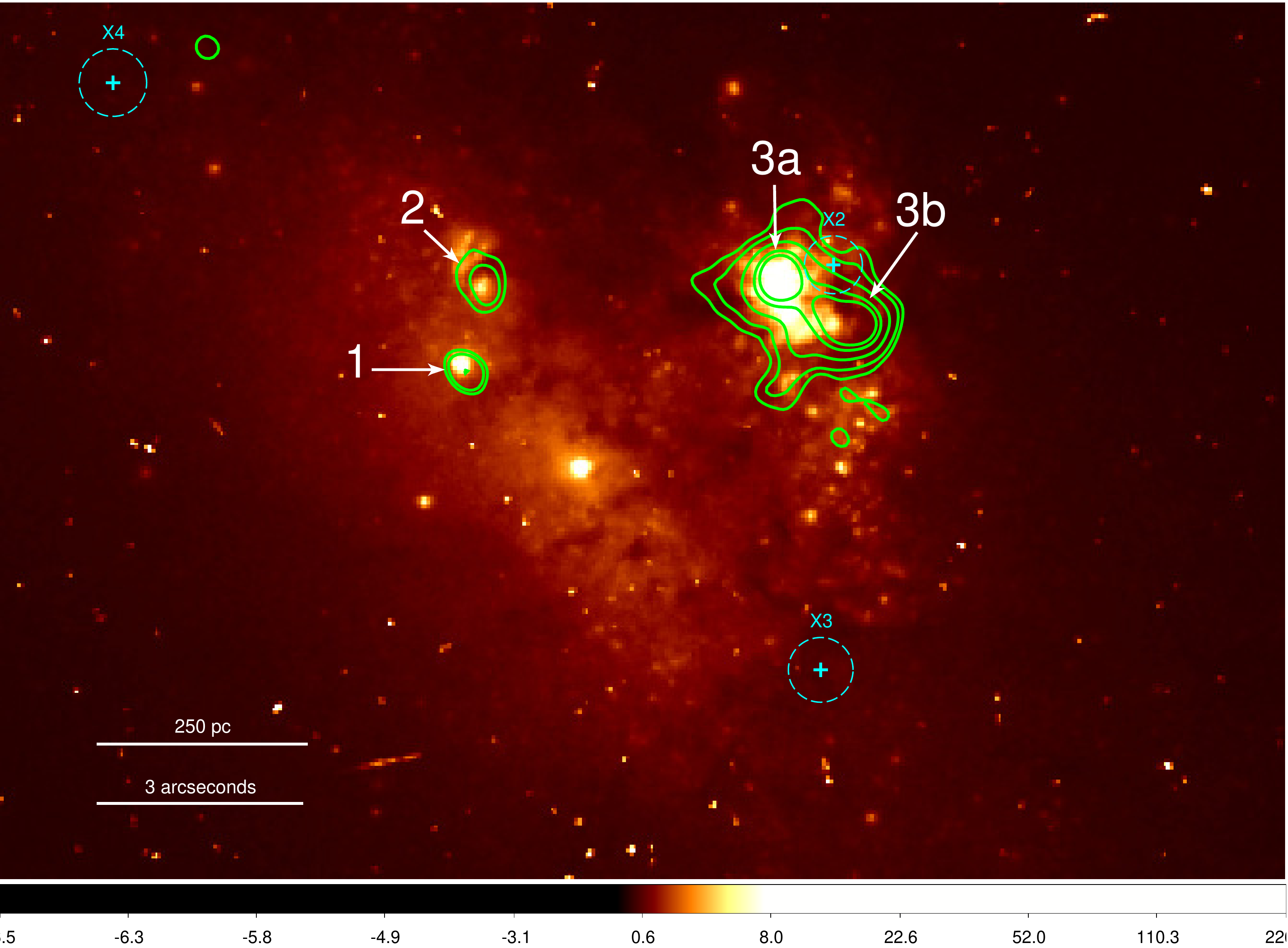}}\hfill
\subfloat[Haro 9 optical image from {\it HST}/WFPC2 (F439W filter). \label{fig:opticalc}]{\includegraphics[width=0.48\textwidth]{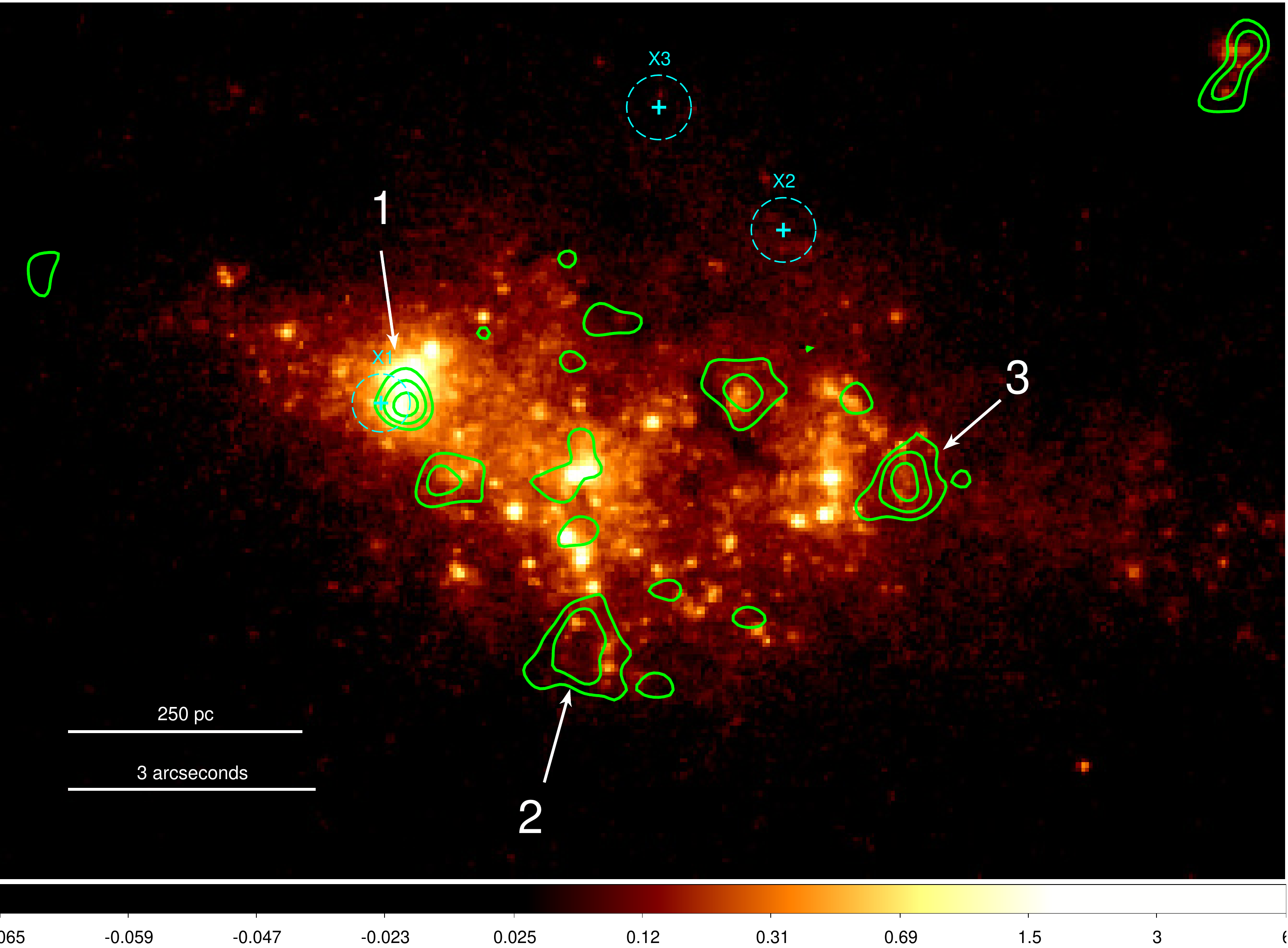}}\hfill
\subfloat[Mrk 996 optical image from {\it HST}/WFPC2 (F569W filter). \label{fig:opticald}]{\includegraphics[width=0.48\textwidth]{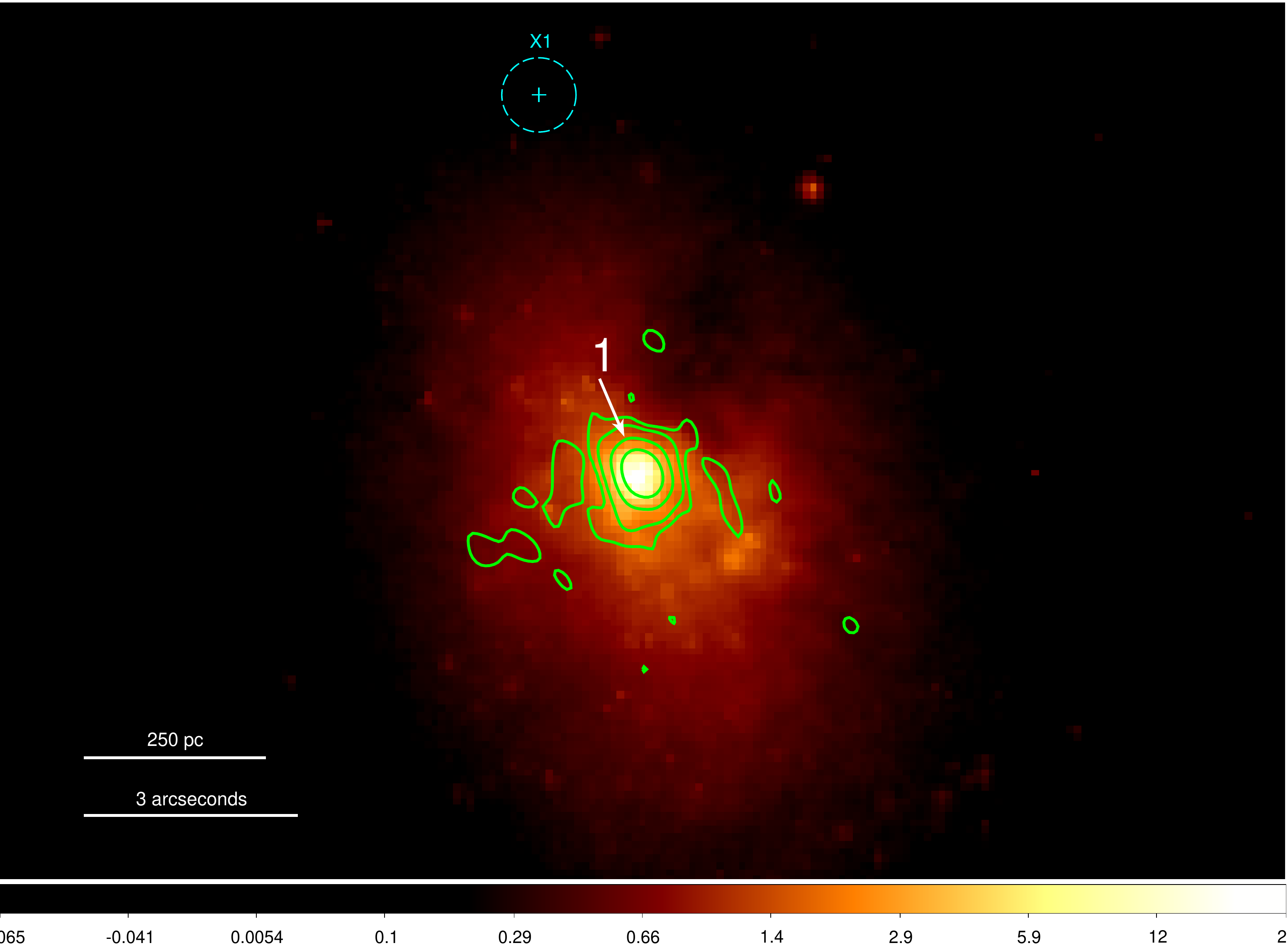}}\hfill
\subfloat[SBS 0940+544 optical image from the SDSS (\textit{g} band). Note that X-ray source X2 is likely a background quasar (see Section \ref{sec:xrayemis}), and thus is excluded from our analysis. \label{fig:opticale}]{\includegraphics[width=0.48\textwidth]{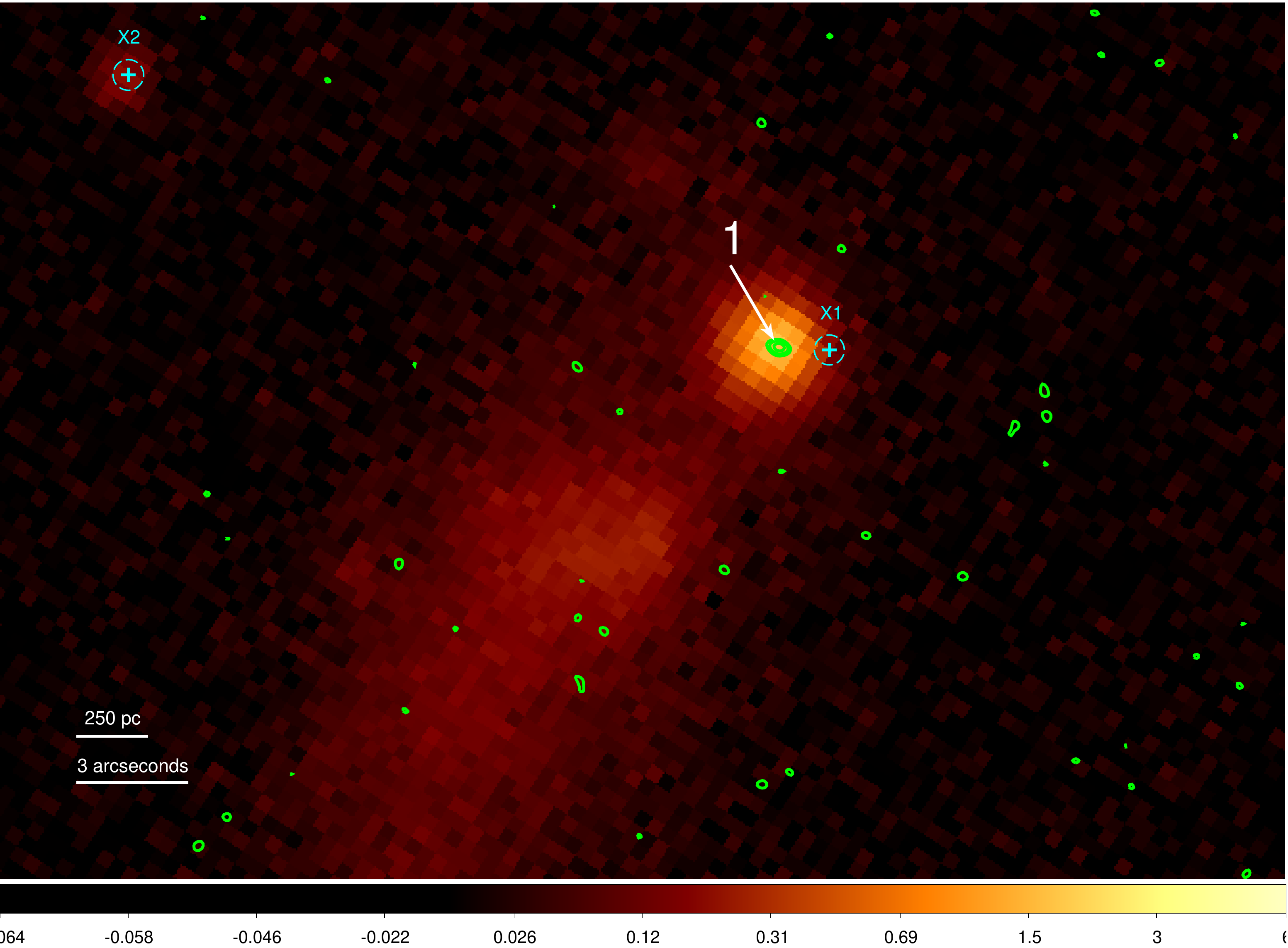}}
\caption{Optical images from {\it HST} or the SDSS as indicated, with X-ray sources (cyan) and 5 GHz radio contours (green) overlaid. For the X-ray sources, the cyan crosses and circles mark the position of the source and the 95\% positional uncertainty. The only case of spatially coincident X-ray and radio sources within the positional uncertainties is in Haro 9, between X1 and R1.} 
\label{fig:optical}
\end{figure*}

As described above, we have detected 10 hard X-ray point sources and 10 compact radio sources in our sample of five blue compact dwarf galaxies.  The luminosities of these sources could be produced by sub-Eddington massive BHs, however the X-ray and radio luminosities alone are also within the expected ranges for stellar processes (see Sections \ref{sec:xrbs}, \ref{sec:radcomp}).  We therefore search for spatially coincident X-ray and radio sources (Figure \ref{fig:optical}), for which the ratio of the luminosities can provide some clues to the origin of the emission.

We find one example of a spatially coincident X-ray and radio source in Haro 9 (X1 and R1 in Figure \ref{fig:opticalc}) within the positional uncertainties.  The hard X-ray source has a luminosity of log $L_{\rm 2-10 keV}=39.4$ and the compact radio source has a luminosity of log $L_{\rm 5 GHz} = 34.7$, where the units are in erg s$^{-1}$.  
We make use of the ratio of the radio and X-ray luminosities following \citet{terashima03},
\begin{equation} \label{eq:rx}
R_{X} = \frac{\nu L_{\nu}(\textrm{5 GHz})}{L_{X} (\textrm{2-10 keV}),}
\end{equation}
and find that log $R_{X} \sim -4.7$.

We see that the source is too luminous in the radio to be a stellar-mass XRB \citep[e.g.,][]{gallo18}, which have log $R_X \lesssim -5.3$. While in principle the radio emission could be due to a radio flare associated with an XRB state transition, with a radio luminosity of log $L_{\rm 5 GHz} \sim 34.7$ the source would be brighter than the brightest flare we have seen from an XRB in our own galaxy \citep[Cyg X-3, log $L_{\rm 15 GHz} \sim 34.3$;][]{corbel12}. Additionally, the likelihood of catching a state transition is rare, especially with both the radio and X-ray emission so bright, given that the observations were taken $\sim$3 weeks apart \citep[state transitions usually last from a few days to a few weeks;][]{yu09}. As such, the source is unlikely to be a stellar-mass XRB, even one undergoing a state transition.

The X-ray/radio source is also not likely to be a SNR, since these objects are typically much more luminous in the radio with log $R_X > -2.7$ \citep[see Supplementary Information in][and references therein]{reines11}.

A more reasonable possibility is an ultra-luminous X-ray source (ULX) bubble. This would account for the X-ray luminosity, and the observed radio luminosity does not exceed values for ULX bubbles found in the literature \citep[e.g.][with log $L_{\rm R} \sim 35.3$]{soria10,cseh12}. While many ULX bubbles are extended, measuring hundreds of pc across in the optical \citep{pakull02,pakull03}, bubbles as small as ${\sim}40$ pc have been observed in the radio \citep{lang07}, which is consistent with the VLA resolution of our observed point-like source. Thus, if the source is a ULX bubble, it would have to be a luminous and relatively compact one.

The source is also consistent with a low-luminosity AGN. From Figure 3 in \cite{merloni03}, we see that the source falls within the range of massive black holes in the compact radio luminosity versus hard X-ray luminosity plane. Under the assumption that the source is indeed an accreting massive BH, the fundamental plane of BH activity relating BH mass to radio and X-ray luminosity implies a BH mass of log~$(M_{\rm BH}/M_\odot) \sim 4.8 \pm 1.1$ using the \citet{merloni03} relation and log~$(M_{\rm BH}/M_\odot) \sim 5.2 \pm 0.44$ using the \citet{millerjones12} relation.  Using the scaling between BH mass and host galaxy total stellar mass (${\sim} 10^{8.38}~M_\odot$ for Haro 9; see Table \ref{tab:sample}) for local AGNs from \citet{reines15} provides a BH mass of log~($M_{\rm BH}/M_\odot) \sim 4.7 \pm 0.55$. If the source is indeed a massive black hole, it must be accreting at a modest rate given the low X-ray luminosity. Assuming a black hole mass of $M_{\rm BH} \sim 10^5$ \msun and an X-ray bolometric correction of 20 \citep{vasudevan09}, the corresponding Eddington fraction would be ${\sim} 0.4\%$.

We note that the ratios and comparisons discussed above are only valid if the X-ray and radio emission are indeed coming from the same source. The positional offset between X1 and R1 in Haro 9 is $0\farcs 3$.  While this is within the positional uncertainties, we cannot definitively say the X-ray and radio emission come from the same object.  Moreover, there is a star cluster complex in the same region as the X-ray/radio source(s) that could plausibly host both an XRB and SNR (or \HII\ region).  Therefore, we caution that this apparently coincident X-ray/radio source in Haro 9 should be considered a {\it candidate} low-luminosity AGN.

\subsection{Detection Limits} \label{sec:detlims}

AGNs in dwarf galaxies are powered by less massive BHs than typical AGNs in more massive galaxies \citep[e.g.,][]{reines15}, and will therefore be less luminous. As an example, the Eddington luminosity of a $10^4 M_\odot$ BH is only $\sim 10^{42}$ erg s$^{-1}$. However, we know that BHs with high accretion rates are rare \citep[e.g.,][]{schulzewisotzki10} so the actual luminosities of AGNs in dwarf galaxies may be orders of magnitude lower.  Therefore, we must lower our luminosity threshold for what {\it could} be an AGN in dwarf galaxy, which we have explored in the preceding sections.  We also require sensitive, high-resolution observations in order to probe lower luminosities and isolate compact X-ray/radio emission from that of the host galaxy on larger scales. 

Here we estimate detection limits of potential massive BHs in our sample of BCDs.  We first estimate the expected BH masses based on the scaling relation between BH mass and host galaxy total stellar mass for local AGNs in \citet{reines15}.  Given the stellar masses in Table \ref{tab:sample}, we expect BH masses in the range log~$(M_{\rm BH}/M_\odot) \sim 3.3 - 4.7$, with a median of log~$(M_{\rm BH}/M_\odot) \sim 4.0$. The scatter in the \citet{reines15} relation is $\sim 0.55$ dex.
Next we use our X-ray detection limits to determine the minimum detectable Eddington fractions of BHs with these masses. The Eddington fraction is given by 
\begin{equation} \label{eq:fedd}
f_{\rm Edd} = (\kappa L_X)/(L_{\rm Edd})
\end{equation}
where $\kappa$ is the 2-10 keV bolometric correction and $L_X$, $L_{\rm Edd}$ are the X-ray and Eddington luminosity of the BH, respectively.
We find $L_{\rm Edd}$ from the BH masses via
\begin{equation} \label{eq:Ledd}
L_{\textrm{Edd}} \approx 1.26 \times 10^{38}~M_{\textrm{BH}}/\textrm{\msun} \textrm{ erg s}^{-1}
\end{equation} 
and take $L_X \sim 10^{38}$ erg s$^{-1}$, which is the minimum detectable X-ray luminosity corresponding to our X-ray flux sensitivity of $F_X \sim 2.5 \times 10^{-15}$ erg s$^{-1}$ cm$^{-2}$ (see Section \ref{sec:xrayemis}) at the distances of our galaxies ($\sim 20$ Mpc).  We assume these BHs are accreting at low rates ($f_{\rm Edd} \lesssim 0.1$) giving $\kappa \sim 15-30$ \citep{vasudevan09}.  Adopting $\kappa \sim 30$ (for an upper bound) and $M_{\rm BH} \sim 10^4~M_\odot$, we find log $f_{\rm Edd} \sim -2.8$.  In other words, our X-ray observations are sensitive enough to detect a $\sim 10^4~M_\odot$ BH radiating at $\sim 0.1\%$ of its Eddington luminosity. The corresponding radio luminosity predicted from the fundamental plane of BH activity \citep{merloni03} is $\sim 10^{33}$ erg s$^{-1}$, albeit with large uncertainty $-$ the $1\sigma$ scatter in the relation is $\sim 0.9$ dex.  Our VLA observations are sensitive to radio luminosities of $\sim$ a few $\times 10^{34}$ erg s$^{-1}$, making this right on the edge of what we could detect (including the $\sim 0.9$ dex uncertainty).

\section{Summary and Conclusions}

We have presented high-resolution \textit{Chandra} and VLA observations of five low-mass blue compact dwarf galaxies in a search for X-ray and radio signatures of accreting massive BHs. We detected 10 hard X-ray point sources and 10 compact radio sources in our sample, which in all but one case did not overlap. While we were primarily interested in detecting signatures of massive BHs, we considered possible alternative origins for the observed radio and X-ray emission such as thermal H{\footnotesize II} regions, SNRs, and XRBs.  None of the X-ray and radio sources alone were so luminous as to rule out star-formation-related emission.  

In the BCD Haro 9, however, we detected a spatially coincident X-ray and radio source (within the astrometric uncertainties) that was consistent with an active massive BH ($M_{\rm BH} \sim 10^5~M_\odot$) under the assumption that the X-ray and radio emission originated from the same source. We cautioned that the combination of a stellar-mass X-ray binary plus a supernova remnant (or \HII\ region) could also account for the X-ray and radio sources, respectively, which are located in the vicinity of a star cluster complex.

Compared to other dwarf galaxies found to host active massive BHs \citep[e.g.,][]{reines11,reines13,reines14}, our sample of BCDs have significantly lower stellar masses ($\sim 10\times$ on average).  Moreover, in contrast to optically-selected samples \citep[e.g.,][]{reines13}, the BCDs studied here are actively star-forming with somewhat disturbed morphologies. The lack of convincing detections of massive BHs in our sample of BCDs could be suggestive of the BH occupation fraction dropping off with decreasing galaxy mass, 
or that BHs in BCDs are not commonly accreting at a detectable rate.  Indeed, recent simulations indicate that BH growth in low-mass galaxies can be stunted by supernova feedback \citep[e.g.,][]{habouzit17,angles17} and that the conditions for rapid BH accretion in dwarf galaxies are rare \citep{bellovary19}.  

While this work has broadened the parameter space in the search for massive BHs in dwarf galaxies, future studies with larger sample sizes are necessary to more accurately determine the prevalence of massive BHs in low-mass star-forming dwarfs such as BCDs and help shed light on the origin of BH seeds \citep[e.g.,][]{plotkinreines2018}.


\acknowledgements
We thank the referee for their comments and gratefully acknowledge helpful conversations from Gregory Sivakoff and Kelsey Johnson during the early stages of this project. Support for this work was provided by NASA through \textit{Chandra} Award Number GO2-13126 issued by the \textit{Chandra X-ray Observatory Center}, which is operated by the Smithsonian Astrophysical Observatory for and on behalf of the NASA under contract NAS8-03060. 
TDR acknowledges support from the Netherlands Organisation for Scientific Research (NWO) Veni Fellowship.
The National Radio Astronomy Observatory is a facility of the National Science Foundation operated under cooperative agreement by Associated Universities, Inc. 
The scientific results reported in this article are based in part on observations made by the \textit{Chandra X-ray Observatory}.
They are also based on observations made with the NASA/ESA Hubble Space Telescope, and obtained from the Hubble Legacy Archive, which is a collaboration between the Space Telescope Science Institute (STScI/NASA), the European Space Agency (ST-ECF/ESAC/ESA) and the Canadian Astronomy Data Centre (CADC/NRC/CSA).
Funding for the Sloan Digital Sky Survey IV has been provided by the Alfred P. Sloan Foundation, the U.S. Department of Energy Office of Science, and the Participating Institutions. SDSS acknowledges support and resources from the Center for High-Performance Computing at the University of Utah. The SDSS web site is www.sdss.org.
This publication makes use of data products from the Wide-field Infrared Survey Explorer, which is a joint project of the University of California, Los Angeles, and the Jet Propulsion Laboratory/California Institute of Technology, funded by the National Aeronautics and Space Administration.
This work has also used observations made with the NASA Galaxy Evolution Explorer. GALEX is operated for NASA by the California Institute of Technology under NASA contract NAS5-98034.
Funding for the NASA-Sloan Atlas has been provided by the NASA Astrophysics Data Analysis Program (08-ADP08-0072) and the NSF (AST-1211644).


\bibliography{ref}

\end{document}